\documentclass[
  aps,
  prl,
  reprint,
  longbibliography,
  superscriptaddress
]{revtex4-2}

\usepackage{amsmath,amssymb}
\usepackage{bm}
\usepackage{graphicx}
\usepackage{xcolor}
\usepackage{hyperref}
\usepackage{physics}
\usepackage{comment}
\usepackage[normalem]{ulem}
\usepackage{xurl}
\usepackage{hyperref}
\usepackage{orcidlink}

\hypersetup{breaklinks=true}

\hypersetup{
  colorlinks=true,
  linkcolor=blue,
  citecolor=blue,
  urlcolor=blue,
  pdftitle={theory_measurement_criticality_tll}
}

\definecolor{myteal}{HTML}{0F6E6C}

\newcommand{\ra}{\rightarrow}

\newcommand{\mc}[1]{\mathcal{#1}}

\newcommand{\bdry}{\mathrm{bdry}}

\newcommand{\unige}{
  Department of Theoretical Physics, University of Geneva,
  24 quai Ernest-Ansermet, 1211 Gen\`eve, Switzerland
}

\newcommand{\gqc}{
  Geneva Quantum Center, University of Geneva
}

\newcommand{\lptms}{
  Universit\'e Paris-Saclay, CNRS, LPTMS, 91405, Orsay, France
}
\usepackage{pdfpages}

\makeatletter
\AtBeginDocument{\let\LS@rot\@undefined}
\makeatother
\begin{document}

\title{Theory of Measurement-Altered Criticality}

\author{Kabir Khanna\:\orcidlink{0000-0001-9400-1597}}
\affiliation{\unige}
 \affiliation{\gqc}

\author{Sara Murciano\:\orcidlink{0000-0002-1638-5692}}
\affiliation{\lptms}

\author{Romain Vasseur\:\orcidlink{0000-0002-4636-4139}}
\affiliation{\unige}
\affiliation{\gqc}

\date{\today}

\begin{abstract}
Local measurements can alter long-range correlations in gapless quantum matter. We propose a theory of weakly-monitored Tomonaga-Luttinger liquids, a broad class of quantum critical states in one dimension. In order to address the intrinsic randomness of the measurement record, we develop a replica instanton calculation to study Born-averaged observables. We find that when measurements are relevant, average correlators of density and phase fluctuations decay at long distances as universal power laws with  logarithmic corrections, a feature we argue is peculiar to measurement-induced randomness. We characterize the full multifractal spectrum of moments of correlations functions, revealing broad, strongly non-gaussian fluctuations across the ensemble of post-measurement states. We support these analytic results with matrix-product-state calculations, and provide a general picture of  measurement-altered criticality for ground states described by 1+1d conformal field theories. Our results establish that physical measurements alter critical quantum states in a manner that lies beyond both forced measurements and conventional critical scaling. 
\end{abstract}

\maketitle
\textit{Introduction --- }Local measurements in quantum mechanics have strikingly non-local effects, as seen for example through the violation of Bell inequalities \cite{PhysicsPhysiqueFizika.1.195, PhysRevLett.49.1804}. While this feature of measurements is well understood in the few-body case, its implications for many-body states remain far less understood. The advent of quantum simulators where such local (even single-site resolved) measurements can be performed has motivated an increasing interest in addressing how local measurements reshape many-body phases of matter, where the central task is to characterize how measurements affect \textit{universal} properties of these phases. In out of equilibrium settings, works have explored measurement induced phase transitions~\cite{PhysRevB.98.205136, PhysRevB.100.134306, PhysRevX.9.031009, PhysRevB.101.104301, PhysRevB.101.104302, hoke_measurement-induced_2023, koh_measurement-induced_2023, annurev:/content/journals/10.1146/annurev-conmatphys-031720-030658, Potter2022} and novel forms of thermalization based on measurements~\cite{PhysRevLett.128.060601, Ippoliti2022solvablemodelofdeep, choi_preparing_2023,PRXQuantum.4.010311, PRXQuantum.4.030322, PhysRevA.107.032215, Chan_2024, PhysRevX.14.041051, 10.21468/SciPostPhys.18.3.107, milekhin2024observable}. In equilibrium settings,  
trivial gapped phases in 1d with exponentially decaying correlations~\cite{PhysRevB.82.155138, Hastings2006}, as intuition suggests, do not exhibit particularly rich measurement-induced phenomena~\cite{PhysRevB.109.195128}. By contrast, topological phases in 1d and 2d display a much richer interplay with measurements, owing to their interesting entanglement structure~\cite{PhysRevB.109.195128}. This has been explored extensively, most notably in the context of measurement-based state preparation~\cite{PhysRevLett.86.5188, PhysRevA.68.022312, briegel_measurement-based_2009, PhysRevLett.119.010504, PhysRevA.96.012302, PhysRevA.98.022332, PhysRevLett.122.090501,daniel2020computational, PhysRevX.14.021040, verresen2021efficiently, PRXQuantum.3.040337, iqbal_non-abelian_2024, foss2023experimental, PhysRevLett.131.200201, chen_nishimori_2025}, where the prepared many-body state serves as a resource for quantum computational tasks.

From the perspective of universality, an especially important class of states is provided by gapless phases governed at low energies by nontrivial fixed points of renormalization-group (RG) flows~\cite{Cardy_1996}. Owing to the absence of a gap, such systems are sensitive to perturbations that couple to low-energy degrees of freedom and can therefore modify their universal long-distance behavior. \textit{Weak} local measurements provide a natural realization of such perturbations: by conditioning the state on local measurement outcomes, they can alter long-range correlations and drive the system towards new measurement-induced regimes. This was first demonstrated in Ref.~\cite{garratt2023measurements} for one-dimensional quantum liquids, namely Tomonaga-Luttinger liquids \cite{tomonaga_remarks_1950, luttinger_exactly_1963, F_D_M_Haldane_1981, giamarchi_quantum_2003}, drawing analogies to the Kane-Fisher impurity problem~\cite{PhysRevLett.68.1220, PhysRevB.46.15233}. Since then, a growing body of work has explored how measurements affect the universal properties of critical states~\cite{PhysRevX.13.041042, PhysRevB.108.165120, PRXQuantum.4.030317, PhysRevB.110.094404, PhysRevB.107.245132, sun2023new,Paviglianiti2024enhanced,tang2024,sala24, patil2024highlycomplexnovelcritical, Lin2023probingsign, khanna2025measurementinducedentanglementconformalfield, khanna2026universalstatisticsmeasurementinducedentanglement, patil2025shannonentropymeasurementrecord, nahum2025bayesiancriticalpointsclassical, 4dwm-kn11,liu24,sarma26,Naus2025, kumar2026universalcrossoversweaklymonitoredquantum}.

\begin{figure}[t!]
    \centering
    \includegraphics[width=\linewidth]{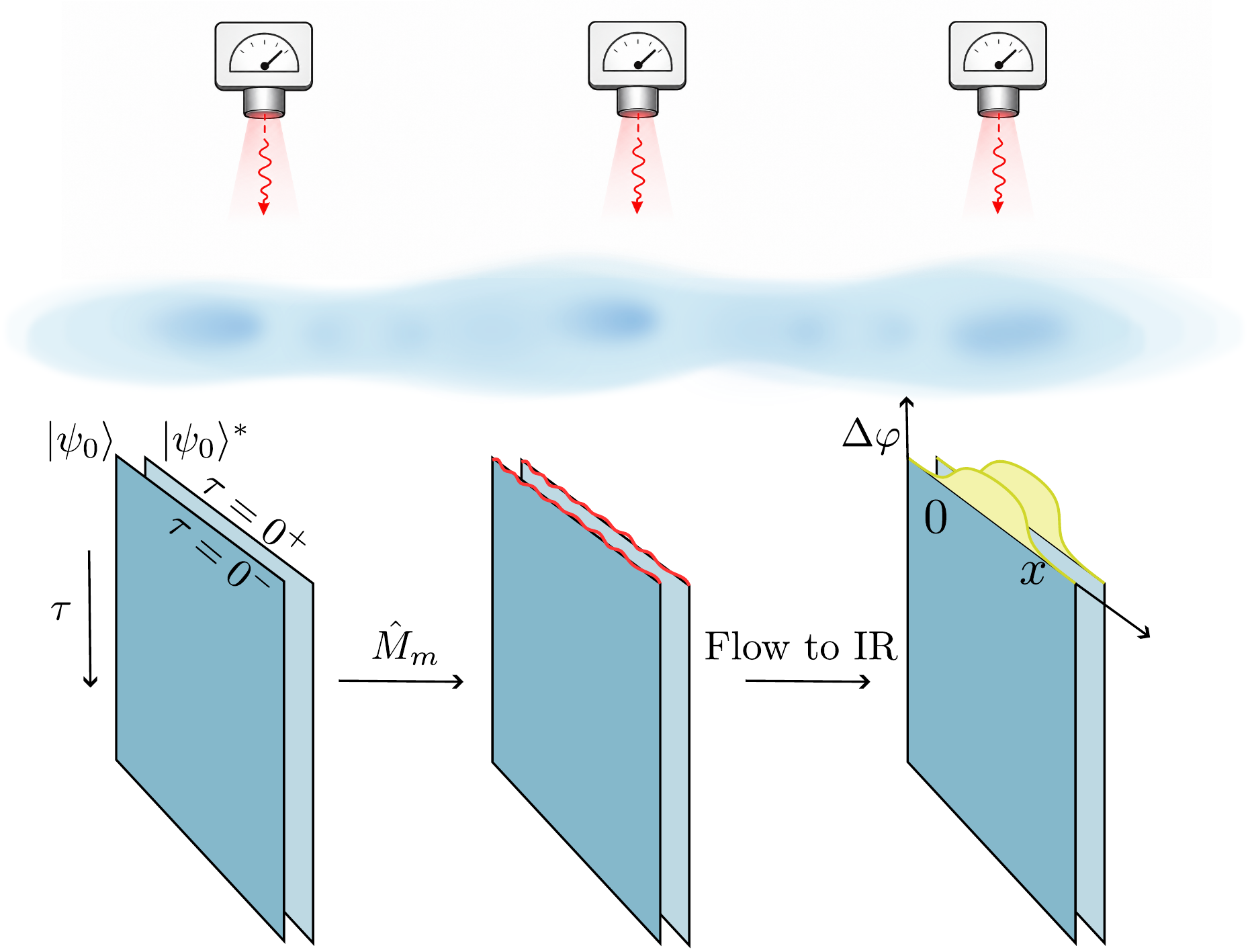}
    \caption{\textbf{Setup.} \textbf{Top:} Schematic illustration of a TLL (blue). Weak measurements are performed on the entire system as shown through the measurement apparatus (red). \textbf{Bottom}: \textbf{Left: } Path integral representation of the ground state density matrix $\ket{\rho_0}\rangle = \ket{\psi_0}\ket{\psi_0^*}$ with the ket and bra defined on the slices $\tau = 0^-$ and $\tau = 0^+$, respectively. \textbf{Middle:} The weak measurement induces a boundary perturbation (red). \textbf{Right:} The replica calculation shows that the averaged long-distance density and phase fluctuations are controlled by field fluctuations $\Delta\varphi$ corresponding to phase-slips between the points 0 and $x$ in the infrared (IR).}
    \label{fig1maintext}
\end{figure} 
A central challenge in this problem, which is also of broader interest, is the intrinsic randomness of measurement outcomes. Most analytical attempts at understanding such measurement-induced phenomena have been restricted to ``forced measurements'', where measurements are treated as deterministic projectors onto favorable configurations. Since this randomness is fundamental to quantum mechanics, any complete characterization of measurement-induced universal behavior must account for the full ensemble of measurement outcomes. A notable exception are the results of Ref.~\cite{patil2024highlycomplexnovelcritical}, which found ``measurement-altered'' fixed points that are perturbatively close to their unmeasured counterparts. In many cases however, the effects of measurements grow in an unbounded way under renormalization. This leaves open the question of how this randomness reshapes the universal long-distance properties in regimes where measurements are weak microscopically,
but the system at long distances is described by strong- or projective-measurement fixed points, requiring non-perturbative treatments. 
\begin{figure*}[t]
    \centering
    \includegraphics[width=\textwidth]{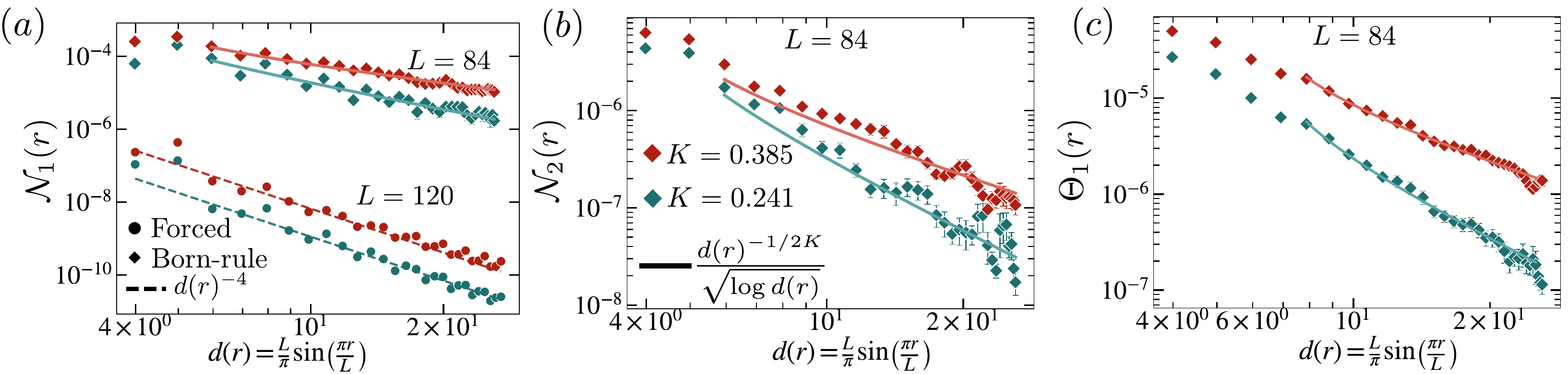}
    \caption{\textbf{Scaling of measurement-altered correlation functions with chord length $d(r)$ for measurement strength $\gamma=0.7$.} Green and red markers correspond to ground states with Luttinger parameters $K=0.241$ and $K=0.385$, respectively. We use $\sum_m p_m\langle \hat S^z_{i_0}\tilde S^z_{i_0+r}\rangle_{m,c}^k$ and $\sum_m p_m \langle\hat S^x_{i_0}\tilde S^x_{i_0+r}\rangle_{m,c}^{2k}$ as microscopic lattice representatives of $\mc N_k(r)$ and $\Theta_k(r)$ respectively~\cite{supplement}, where $\tilde S^{\alpha}_i = 3^{-1}(\hat S^\alpha_{i-1} + \hat S^\alpha_{i} + \hat S^\alpha_{i+1}$), $\hat S^{\alpha=x,y,z}_i$ are Pauli spin-1/2 operators, and $\langle\dots\rangle_{m,c}$ denote connected correlators for measurement outcome $m$. Solid lines of the corresponding colors show fits to the theoretical predictions in Eq.~\eqref{densitycorrscaling}, while the dotted line denotes the prediction for the correlator after a forced measurement that post-selects a CDW state, made in~\cite{garratt2023measurements}. (a) Born-averaged connected density correlator $\mc N_1(r)$ compared with the forced correlator. (b) Second moment of the Born-averaged connected density correlator, $\mc N_2(r)$. (c) Born-averaged phase correlator $\Theta_1(r)$.}
    \label{corrplot}
\end{figure*}

In this Letter, we address this problem for a broad class of 1d quantum liquids, and provide a general picture that we expect to hold for a class of such strong-measurement fixed points. 
%We develop a replica framework, allowing us to use the replica trick to treat the randomness of measurement, drawing inspiration from Refs.~\cite{PhysRevB.101.104301, PhysRevB.101.104302,garratt2023measurements} where such a trick was first applied to study measurement-induced disorder. 
%We develop a replica framework that allows us to treat the randomness of measurement via the replica trick, \kk{["drawing inspiration from"? ] instead of "building in particular on"}building in particular on Ref.~\cite{garratt2023measurements}, where this approach was introduced to study random outcomes in measurement-altered criticality. 
We develop a replica framework that allows us to treat the randomness of measurement via the replica trick, drawing on the replica approach of Ref.~\cite{garratt2023measurements}, where it was introduced in the context of measurement-altered criticality. We show that the resulting universal behavior can differ sharply from that obtained by conditioning on special measurement outcomes, a settings we refer to as \textit{forced} measurements, in contrast to \textit{real} measurements where one must account for the entire ensemble of post-measurement states. 

\textit{Setup---} The class of systems we consider in this work are 1d Tomonaga Luttinger liquids (TLLs), which describe a host of systems such as interacting bosons, fermions, and spin chains in 1d, at low energies. For simplicity, we consider \textit{spinless} quantum liquids governed by the Euclidean action \cite{giamarchi_quantum_2003}
\begin{equation}
\label{tllaction}
    S[\varphi] = \frac{1}{2\pi K} \int_0^L dx\int_{0}^{\beta} d\tau \left[(\partial_x\varphi)^2 + (\partial_\tau\varphi)^2\right],
\end{equation}
where $K$ is the Luttinger parameter, $x$ and $\tau$ are space and imaginary time coordinates, $L$ is the size of the system, $\beta$ the inverse temperature, and where the field obeys $\varphi(x,0) = \varphi(x,\beta)$. The physical meaning of the field $\varphi(x,\tau)$ is most simply understood from its relation to the particle density 
$n(x,\tau) = -\pi^{-1}\partial_x\varphi(x,\tau) + \pi^{-1} \cos[2k_F x - 2\varphi(x,\tau)] + \dots$, with $k_F$ being the Fermi wave-vector fixed by the filling fraction.
A key property of the bosonic field $\varphi$ is its compactness: $\varphi(x,\tau) \sim \varphi(x,\tau) + \pi$, as reflected in the invariance of the leading $2k_F$ oscillatory component.
In addition to density fluctuations that are captured via $\varphi(x,\tau)$, the canonically conjugate momentum field $\pi \Pi(x,\tau) = \partial_x\theta(x,\tau)$ describes phase fluctuations and obeys $[\varphi(x), \partial_{x'}\theta(x')] = i\pi \delta(x-x')$. In this work, we focus on the ground state of this model which exhibits algebraically decaying correlations with exponents set by $K$. 

Formally, the ground state is obtained through imaginary time evolution as $\ket{\psi_0}\bra{\psi_0} \propto \lim_{\beta\ra\infty} e^{-\beta \hat H}$, where the Hamiltonian $\hat H$ obeys
$Z = \Tr e^{-\beta \hat H} = \int D\varphi\; e^{-S[\varphi]}$. We now consider weak measurements of an observable $\hat O(x)$ on $\ket{\psi_0}$ with a measurement strength $\gamma \in [0,\infty)$, where the limit $\gamma\rightarrow \infty$ corresponds to projective measurements. Such operations can be implemented by the Kraus operator $\hat M_m =\exp[- \frac{\gamma}{4} \int dx [\hat O(x) - m(x)]^2]$, where $m(x)$ labels the measurement outcome. In the Euclidean path integral this corresponds to a perturbation localized at the \textit{boundary} $\tau=0$, where $\tau\in(0,\beta)$ with $\beta \ra\infty$.
The measurement outcomes define an ensemble of normalized states $\ket{\psi_m}=\hat M_m\ket{\psi_0}/\sqrt{p_m}$, with Born probabilities $p_m\equiv Z_m/Z=Z^{-1}\int D\varphi \exp[-S[\varphi]-\frac{\gamma}{2}\int_{\tau=0} dx (O(x)-m(x))^2]$, with $\hat O(x)$ a function of $\hat\varphi(x)$, so it is diagonal in the field basis. The normalization is fixed by $\sum_m p_m=\sum_m Z_m/Z=1$. Our goal is to characterize the universal long-distance properties of this measurement-induced ensemble. In this work, we focus on the distribution of the following connected correlators:
\begin{align}
\label{densitycorr}
    \mathcal N_k(x) &= \sum_m p_m [\langle  \hat n(x)  \hat n(0)\rangle_m -\langle  \hat n(x)\rangle_m\langle  \hat n(0)\rangle_m]^k,\\
\label{phasecorr}
    \Theta_k(x) & = \sum_m p_m |\langle e^{i[\theta(x) - \theta(0)]}\rangle|_m^{2k},
\end{align}
where $\langle \dots\rangle_m$ denotes an expectation value w.r.t $\ket{\psi_m}$. In what follows, we fix $\hat O(x) = \hat n(x)$, that is, we monitor the local density of this model. Ref.~\cite{garratt2023measurements} first studied this setup post-selecting on a charge-density-wave (CDW) outcome: for $K<1$ the forced measurement is relevant, so arbitrarily weak measurements pin $\varphi$ to a saddle-point minimum $w\pi$, $w\in\mathbb Z$, altering long-distance correlations. Compactness of $\varphi$, however, permits phase slips connecting minima differing by $\pi$; these leading irrelevant defects give rise to a dilute Coulomb gas that governs the long-distance density and phase correlations. As we will see, this mechanism persists in measurement-averaged quantities, where one must additionally account for the randomness of outcomes.

\textit{Replica Field Theory---} We use the replica trick to treat the randomness of measurement outcomes~\cite{PhysRevB.101.104301, PhysRevB.101.104302}. Introducing $Q$ replicas coupled through the shared measurement record $m$, measurement-averaged correlators become ordinary correlators in the replicated theory; the desired Born average is recovered by analytically continuing $Q$ and taking $Q\to1$. For example, the mean can be written through the replica construction as $\mc N_1(x)=\lim_{Q\to1}\langle n_1(x)n_1(0)-n_1(x)n_2(0)\rangle_{S_Q}$, where~\cite{garratt2023measurements}
\begin{equation}
\label{effectiverepaction}
S_Q = \sum_{a=1}^Q S[\varphi_a]
-\frac{\gamma}{2Q\pi^2}\int dx\sum_{b<c}
\cos \left. \left[2(\varphi_b-\varphi_c)\right] \right|_{\tau=0},
\end{equation}
is the effective action obtained after integrating over measurement outcomes, with irrelevant terms omitted. The replica perturbation stems from the oscillatory $\cos(2k_Fx-2\varphi)$ term in the density $n(x)$, where $\varphi(x)$ acts as a phase of the cosine. Consequently, the measurement suppresses \textit{relative} phase fluctuations between replicas. The scaling dimension of the replica perturbation is $2K$, so it is relevant for $K<1/2$~\cite{garratt2023measurements}, corresponding to strongly repulsive fermions. We henceforth focus on this regime, where $\gamma$ flows to the strong-coupling fixed point $\gamma\to\infty$. The detailed solution of the replicated field theory in this limit is given in the supplemental material~\cite{supplement}; here we summarize the main ideas.

The replica potential in Eq.~\eqref{effectiverepaction} is invariant under common shifts $\varphi_a\to\varphi_a+\delta\varphi(x)$. We therefore rotate to relative and center-of-mass (c.o.m) modes, following Fradkin and Moore~\cite{PhysRevLett.97.050404}, by defining $(\tilde{\bm\varphi},\tilde\varphi_Q)^t=U_Q\bm\varphi$. Here $\tilde{\bm\varphi}=(\tilde\varphi_1,\ldots,\tilde\varphi_{Q-1})$ are the relative fields, while $\tilde\varphi_Q$ is the c.o.m field. In this basis the Gaussian bulk action is unchanged, while the boundary perturbation acts only on the relative fields. At the $\gamma\to\infty$ fixed point, $\tilde{\bm\varphi}$ is uniformly pinned at the boundary to one of the minima $\tilde{\bm\varphi}=\pi U_{Q-1}\bm w$, with ``charge" $\bm w\in\mathbb Z^{Q-1}$ and $U_{Q-1}$ the upper-left $(Q-1)\times(Q-1)$ block of $U_Q$. The c.o.m mode decouples and remains free at the boundary.

The correlators in Eqs.~\eqref{densitycorr} and \eqref{phasecorr} can then be evaluated directly at this fixed point. We refer to the corresponding universal exponents as \textit{boundary conformal field theory} (BCFT) contributions, since they arise from conformally invariant (uniform) boundary conditions on $\bm{\tilde\varphi}$. Leading irrelevant perturbations around this fixed point lead to instanton solutions, or phase slips, which are tunnelings between different pinned solutions that shift the charge $\bm w\to\bm w+\Delta\bm w$. Although irrelevant, we will see that these perturbations are crucial because they can produce a slower universal decay of correlations than the fixed-point BCFT contribution.

Phase slips are incorporated perturbatively about the strong-coupling fixed point, where the expansion is controlled by the fugacity $g(\Delta\bm w)$ of instantons that shift charge by $\Delta\bm w$. For two-point functions of interest in this work, the leading non-trivial contribution appears at $\mc O(g^2)$ and consists of a neutral pair of slips of charges $\Delta\bm w,-\Delta\bm w$ inserted at points $0$ and $x$ where the correlators are evaluated. The resulting effective theory in this 2-slip sector is a multi-component Coulomb gas with an effective interaction $S_{\bdry, Q} = -\frac{2}{K}\Delta\bm w T_{Q-1}\Delta\bm w\log|x|$, where $T_{Q-1} = U_{Q-1}^TU_{Q-1}$ sets the inter-replica interaction. One must then evaluate the correlators in this multi-component Coulomb gas, analytically continue the result, and take the replica limit $Q \ra 1$. Below we describe the resulting expression for the average density correlators; the corresponding derivation and expression for the phase correlator is in the supplementary material~\cite{supplement}.
In the replica limit $Q \ra 1$, we find
\begin{equation} 
\label{densityexpression}
\mc N_k(x) = \int_0^{\pi}\frac{d\delta}{\pi}\frac{e^{-F_x(\delta,0)}}{\mc W} \left[-\partial_J^2 F_x(\delta,J)\vert_{J=0}\right]^k,
\end{equation}
where $F_{x}(\delta, J)= -\log\left[\mathcal{ Z}_x(\delta,J)\right]$ and $\mathcal Z_x(\delta,J) = \sum_{w\in\mathbb Z} x^{-\frac{2}{K}\left(w+\frac{\delta}{\pi}\right)^2 + J\left(w+\frac{\delta}{\pi}\right)}$ is a partition function for the slips labeled by an offset $\delta\in(0,\pi)$, with a source $J$ that couples to the phase-slip charge $w+\delta/\pi$. $\delta$ is best interpreted as the reminiscent $\delta\varphi(x)$ phase degree of freedom that is not fixed by the measurement and hence enumerates the set of effective measurement outcomes in the IR. The smooth density $n(x) \sim -\partial_x\varphi\propto (w+\delta/\pi)$  measures the \textit{charge} of the phase slips of the same magnitude $w+\delta/\pi$ at $0$ and $x$. Thus the connected correlator $\langle n(x)n(0)\rangle-\langle n(x)\rangle\langle n(0)\rangle$ probes the variance of this phase-slip charge in the fixed $\delta$ sector, encoded by $\partial_J^2F_x(\delta,J)|_{J=0}$, with $\mathcal Z_x(\delta,0)$ providing the corresponding background weight. Finally, one must average over $\delta$ (coarsed-grained effective measurement outcomes) with the normalized ``Born" weights given by $\mc W^{-1}\mathcal Z_x(\delta,0)$ with $\mc W^2 = 2\log x/(K\pi) $ so that $\int_0^\pi  \frac{d \delta}{\pi}  \mc  W^{-1}\mathcal Z_x(\delta,0) = 1$.

\textit{Scaling of Correlations---}The leading universal decay of averaged correlators \eqref{densitycorr} and \eqref{phasecorr} is $\mc C_k \sim \mathcal M_k(x) + x^{-\Delta_{\mc C}(k)}$ with $\mc C_k(x)=\mc N_k(x),\Theta_k(x)$, and where
\begin{equation}
\label{densitycorrscaling}
\mathcal M_k(x) \sim\begin{cases}
     x^{-\frac{2k(1-k)}{K}},\,\, &0< k\leq 1/2 \\
    x^{-\frac{1}{2K}}/\sqrt{\log x},\,\, &k>1/2
\end{cases},
\end{equation}
is the multifractal~\cite{LUDWIG1990639, PhysRevLett.66.247} contribution due to averaging over the effective measurement outcomes $\delta$. In contrast, $\Delta_\Theta(k) = 2k/K $ and $\Delta_{\mc N}(k) = 4k$ are saddle-point (``BCFT'') contributions that are present for all measurement outcomes. These outcome-independent terms are always sub-leading for $\Theta_k(x)$ but can become the leading contribution for $\mc N_k(x)$ when $K$ is below a certain threshold.

In the clean TLL, the (smooth part of) density and phase correlations decay as $\langle n(x)n(0)\rangle_0\sim x^{-2}$ and $\langle e^{i[\theta(x)-\theta(0)]}\rangle_0\sim x^{-1/(2K)}$, respectively. Under forced density measurement, Ref.~\cite{garratt2023measurements} found a faster decay for density correlators, $\sim x^{-4}$, together with phase correlations scaling as $\sim x^{-1/K}$ for $K<1/2$. The intuitive reason is that density measurements suppress long-range density fluctuations and correspondingly enhance phase correlations.
However, in the measurement-averaged problem, both correlators exhibit the same scaling (for $k>1/2$), and their asymptotic decay differs from that found in both the clean and forced-measurement cases.
This atypical behavior can be understood as follows. The Born weight is concentrated around typical phase shifts, while the probability density near $\delta=\pi/2$ is algebraically suppressed as $x^{-1/(2K)}$. Nevertheless, at $\delta=\pi/2$, the $w=0$ and $w=-1$ sectors are degenerate, so both density and phase fluctuations remain of order one, while they decay algebraically for generic $\delta$. As a result, the narrow, low-probability region around $\delta=\pi/2$ controls the leading decay of all moments of both correlators.
A further peculiarity of the measurement-averaged correlators is the multiplicative factor $(\log x)^{-1/2}$. This factor arises from integrating over a continuum of phase offsets in a narrow region around $\delta=\pi/2$, where the $w=0$ and $w=-1$ sectors remain nearly degenerate. Such a multiplicative correction is a \textit{universal} characteristic of the strong-coupling measurement-averaged problem, and distinguishes it from the pure scale-invariant power laws familiar in the weak measurement fixed points  found in~\cite{patil2024highlycomplexnovelcritical}. Similar logarithmic terms were also observed for measurement induced entanglement (MIE)~\cite{khanna2025measurementinducedentanglementconformalfield, khanna2026universalstatisticsmeasurementinducedentanglement}. 

\textit{Numerics--- }We support our results with numerical benchmarks on a lattice model with spin-1/2 degrees of freedom with the Hamiltonian 
\begin{equation}
    H = \sum_i -t(S^+_i S^-_{i+1} + S^-_i S^+_{i+1}) + U_1 S^z_i S^z_{i+1} + U_2 S^z_i S^z_{i+2},
\end{equation}
where $S_i^{\pm} = S_i^x\pm iS_i^y$. This model reduces to non-interacting fermions with $K = 1$ when $U_1 = U_2 = 0$~\cite{giamarchi_quantum_2003}. Away from this limit, fixing filling fraction to be $n_0 = 1/3$, one can access a Luttinger liquid with $K<1/2$ for a range of parameters with $U_1,U_2>0$ \cite{{PhysRevResearch.3.013114, PhysRevResearch.4.043225}}. For the results in this work, we use the iTensor library~\cite{itensor} to obtain approximate ground states for the parameters $U_1 = 10$, and $U_2 =1.0, 2.5$. Since there is no available non-perturbative relationship between $K$ and the parameters of the model, we infer the value of $K$ from the known algebraic decay of correlation functions in a Luttinger liquid~\cite{giamarchi_quantum_2003}. For the results in this work, we use two values of $K\approx 0.384, 0.241$ obtained using the parameter values $U_1=10, U_2 = 1.0, 2.5$ respectively~\cite{supplement}. Upon obtaining the approximate ground states, we perform weak measurements in the $S^z$ basis using the measurement operator $\hat M_m =\prod_i \hat M_{m_i} = \prod_i e^{2\gamma m_iS^z_i}$, where $\gamma$ is the measurement strength and $m = (m_1,\dots m_L)$ is the set of measurement outcomes with $m_i\in\{\pm 1\}$. The measurements are sampled as per the Born rule, after which we evaluate the required correlation functions and average over measurement trajectories. For the forced case, we post-select to the CDW pattern $\ket{\uparrow\downarrow\downarrow\uparrow\downarrow\downarrow\dots}$ for $n=1/3$. Our numerical results are shown in Fig.~\ref{corrplot}. As seen in the plots, fitting the non-universal factors in Eq.~\eqref{densitycorrscaling} captures the observed scalings sufficiently well. We provide additional numerical checks such as convergence with system size and measurement strength in the supplementary material~\cite{supplement}.

\textit{General Picture --- }    The results above fit naturally into a conjectured general framework based on BCFT \cite{CARDY1984514, cardy2004boundary, CARDY1989581, PhysRevLett.67.161, AFFLECK1991641} for ground states governed by CFTs. In the doubled state representation of the ground-state density matrix, $\ket{\rho_0}\rangle=\ket{\psi_0}\ket{\psi_0^*}$, the bra and ket live on the two $\tau=0_\pm$ boundaries of a Euclidean manifold. Measurements act as boundary perturbation of the \textit{doubled} CFT along the imaginary time slice $\tau = 0$, where both copies are constrained to have the same measurement outcomes (see Fig.~\ref{fig1maintext}). When relevant, this perturbation drives a boundary RG flow towards an ensemble of \textit{conformally invariant boundary conditions} (CIBCs) of the doubled theory. In principle, the ensemble of CIBCs reached depend on coarse-grained properties of the measurement, such as the symmetries it preserves. However, finding this ensemble can be a hard task in general, requiring an understanding of the underlying (boundary) RG flow. Besides CIBCs, a boundary CFT is also characterized by a spectrum of boundary condition changing (BCC) operators that change boundary condition between one CIBC ($a$) to another $(b)$ when inserted at a point. These operators and their respective scaling dimensions are enumerated in the boundary-partition functions of the respective BCFT on an annulus with two CIBCs $a$ and $b$. 

Once the aforementioned boundary RG flow is characterized and the BCFT data are known, we posit that measurement-averaged observables can be calculated by averaging the observables evaluated in the ensemble of CIBCs with possible BCC inserted. The Born-weights associated with this average are given by the associated partition function of the boundary with possible BCC insertions. One can make this picture concrete in the case of the TLL considered in this work, where the underlying CFT is the free boson CFT. While the details of this are in the supplementary material~\cite{supplement}, the above idea remains. Local density measurements drive the boundary to an ensemble of conformally invariant Dirichlet boundary conditions (DBC) labeled by $\delta\in[0,\pi)$. The associated BCCs that change boundary condition from $\delta_1\ra \delta_2$ are $\hat{\mathcal O}_{\delta,w}={e^{i(w+\delta/\pi)\theta}}$ such that $\delta = \delta_1 - \delta_2$ (mod $\pi$), with scaling dimensions $\Delta_{\delta,w} =(2K)^{-1}(w+\delta/\pi)^2$ and $w\in\mathbb Z$. The partition function $\mc Z_x(\delta,0)$ enumerates these BCC channels and gives the Born weight for a fixed $\delta$. Measurement-averaged correlators are then given by evaluating the correlators for a fixed $\delta$ and averaging over $\delta$ with the Born weight proportional to $\mc Z_x(\delta,0)$, such as in Eq.~\eqref{densityexpression}. 

\textit{Summary --- } We have developed a replica-based framework to determine the IR fate of Born-averaged correlation functions upon weakly measuring the density in a class critical 1d quantum states: Tomonaga Luttinger liquids. Using this framework, we find that the resulting moments of measurement averaged correlation functions of density and phase share the same long-distance scaling Eq.~\eqref{densitycorrscaling} for $k>1/2$, in sharp contrast to the case where one post-selects to the CDW state \cite{garratt2023measurements}. The resulting scaling is strongly non-gaussian in the distribution of post-measurement correlations and is governed by a
narrow range of phase offset with anomalously large density and phase correlations.

We interpret our results in a more general BCFT framework that we expect to generalize to other critical ground states whose low-energy properties are captured by CFTs, such as the Ising model. A natural continuation of our work is therefore to test how generally this BCFT pictures holds for other CFTs. We anticipate this to be challenging as it requires a detailed understanding of random line defects and conformally invariant boundary conditions in doubled CFTs with central charge $c \geq 1$~\cite{OSHIKAWA1997533, Jeng_2001}. In the Ising CFT, the doubled-theory construction and the associated conformal boundary conditions have already been studied for some line defects~\cite{OSHIKAWA1997533, Jeng_2001}. Nevertheless, extending this analysis to more general defects and measurement protocols remains nontrivial. 
More generally, one must determine the relevant boundary conditions and the operator content in the doubled CFT. Since the double theory has central charge $c\geq 1$, its spectrum may become richer and can contain continuous families of conformal boundary conditions and boundary operators. Identifying the boundary fixed points with the corresponding boundary-condition changing operators is therefore expected to be challenging. Finally, it would be interesting to probe such altered-critical correlations in platforms such as Rydberg atom arrays where concrete proposals for doing so already exist in some cases \cite{Naus2025}. 

{\it Data availability --} The data and codes used to generate Fig.~\ref{corrplot} and Figs.~1-4 in the supplementary material \cite{supplement} are available on
Zenodo~\cite{khanna_2026_21791465}.

 {\it Acknowledgments -- } This work was supported by the Swiss National Science Foundation (grant 10008234) and by the Foundation for the University of Geneva. 
 We thank Ehud Altman, Gabriel Cuomo, Sam Garratt, Sarang Gopalakrishnan, Abhishek Kumar, Andreas Ludwig, Marco Meineri, Rushikesh Patil, and Pablo Sala, for insightful discussions and/or collaborations on related topics. The authors also acknowledge the use of Codex and ChatGPT 5.6 (OpenAI) for assistance with implementing and debugging numerical calculations used in this work and in processing and plotting the resulting numerical data. All resulting code and calculations were independently checked and validated by the authors.

% \bibliographystyle{apsrev4-2}
% \bibliography{apssamp}

\bigskip
\onecolumngrid
\newpage


\section*{Supplementary material}

# Supplemental Material

## Theory of Measurement-Altered Criticality

Kabir Khanna[1,2], Sara Murciano[3], and Romain Vasseur[1,2]
[1] *Department of Theoretical Physics, University of Geneva, 24 quai Ernest-Ansermet, 1211 Genève, Switzerland*
[2] *Geneva Quantum Center, University of Geneva*
[3] *Université Paris-Saclay, CNRS, LPTMS, 91405, Orsay, France*

### CONTENTS



### I. REPLICA FIELD THEORY

Here we present a detailed derivation of the replica field theory used to evaluate correlators after the measurement of local density on Tomonaga Luttinger Liquids (TLLs) described by the Euclidean action [1]

$$S[\varphi] = \frac{1}{2\pi K}\int dx \int_0^\beta d\tau[(\partial_x\varphi)^2 + (\partial_\tau\varphi)^2], \tag{1}$$

where the field is compactified with $\varphi \sim \varphi + \pi$. The Hamiltonian $\hat{H}$ is related to the action via $\mathrm{Tr}\, e^{-\beta\hat{H}} = \int D\varphi e^{-S[\varphi]}$ with $\beta^{-1}$ setting the temperature. We are interested in *weakly* monitoring the local density in the ground state $|\psi_0\rangle$ of this model, where the local density is written as [1]

$$\hat{n}(x) = -\frac{1}{\pi}\partial_x\hat{\varphi} + \frac{A}{\pi}\cos(2k_F x - 2\hat{\varphi}) + \dots, \tag{2}$$

with $A$ a nonuniversal prefactor that we will set to unity for simplicity as it can be absorbed in the measurement strength. The operator that implements a weak measurement is

$$\hat{M}_m \propto \exp\left[-\frac{\gamma}{4}\int dx\, (m(x) - \hat{n}(x))^2\right], \tag{3}$$

where $\gamma$ is the measurement strength and the pre-factor is determined by requiring the completeness condition $\sum_m \hat{M}_m^2 = 1$. One can interpret the above equation as a “weak” measurement by thinking of $m(x)$ as being a

Gaussian variable spread around the true value $n(x)$ of density: the larger the coefficient $\gamma$, the smaller the uncertainty in the measurement. Experimentally, such an operation can be implemented by uniformly coupling each site of the primary chain to that of an ancilla chain and performing usual on-site projective measurements on the ancilla chain. By tuning the coupling between the ancilla and the primary chain, one can realize the whole range of measurement strengths $\gamma \in [0, \infty)$ (see Refs. [2, 3] for concrete protocols). The post-measurement state $|\psi_m\rangle$ with measurement outcome $m$ is written as

$$|\psi_m\rangle = \frac{\hat{M}_m |\psi_0\rangle}{\sqrt{p_m}} \quad , \quad p_m = \langle \psi_0 | \hat{M}_m^2 | \psi_0 \rangle, \tag{4}$$

where $p_m$ is the Born probability for the outcome $m$. In the Euclidean path integral representation, one can write

$$p_m = \lim_{\beta \to \infty} \frac{\mathrm{Tr}\left[e^{-\beta \hat{H}} \hat{M}_m^2\right]}{\mathrm{Tr}\left[e^{-\beta \hat{H}}\right]} = \frac{Z_m}{Z} = \int D\varphi \left[\hat{M}_m^2 \exp\left[-S[\varphi]\right]\right], \tag{5}$$

where we set $Z = 1$ hereon for simplicity.

### A. Replica Action and Saddle Point

We are interested in the long-distance scaling of averaged density and phase correlators $\mathcal{N}_k(x)$ and $\Theta_k(x)$. For quantities that are *linear* in the density matrix $|\psi_m\rangle \langle\psi_m|$, such averages, generally speaking, do not lead to measurement altered correlations. This follows from the re-writing of the measurement-average $\sum_m p_m \langle \hat{O} \rangle_m$ for any observable $\hat{O}$ as the ground state expectation of the observable de-phased in the measurement basis, that is, $\sum_m p_m \langle O \rangle_m = \mathrm{Tr}[\rho_0 \mathcal{E}_M(O)]$, where $\rho_0 = |\psi_0\rangle \langle\psi_0|$ is the ground state density matrix and $\mathcal{E}_M(O) = \sum_m \hat{M}_m O \hat{M}_m$. For $\hat{O} = \hat{n}$, $[\hat{O}, \hat{M}] = 0$ and so the action of this channel is trivial due to $\sum_m \hat{M}_m^2 = 1$, resulting in the measurement averaged correlator being unaltered. For $\hat{O} = e^{i\hat{\theta}(x)}$, one can check from the action of the measurement channel (using Eq. (3)) that it keeps $e^{i\theta}$ unchanged up to a non-universal factor that depend on measurement strength. Therefore, for finite measurement strengths $\gamma$, the measurement-averaged correlator is not altered (up to a non-universal pre-factor) in this case too.

In contrast, nonlinear quantities such as the ones considered in the main text can indeed display altered-correlations and need a replica trick to evaluate them. For instance, $\mathcal{N}_1(x)$ in the replicated formalism can be expressed as

$$\mathcal{N}_1(x) = \sum_m p_m [\langle \hat{n}(x)\hat{n}(0)\rangle_m - \langle \hat{n}(x)\rangle_m \langle \hat{n}(0)\rangle_m] \tag{6}$$

$$= \lim_{Q\to 1} \frac{\sum_m p_m^Q [\langle \hat{n}(x)\hat{n}(0)\rangle_m - \langle \hat{n}(x)\rangle_m \langle \hat{n}(0)\rangle_m]}{\sum_m p_m^Q} \tag{7}$$

$$= \lim_{Q\to 1} \frac{\int D\boldsymbol{\varphi} [n_1(x)n_1(0) - n_1(x)n_2(0)] \int Dm(x) \left[\prod_{a=1}^Q \hat{M}_{m,a}^2 \exp\left[-S[\boldsymbol{\varphi}]\right]\right]}{\int D\boldsymbol{\varphi} \int Dm(x) \left[\prod_{a=1}^Q \hat{M}_{m,a}^2 \exp\left[-S[\boldsymbol{\varphi}]\right]\right]}, \tag{8}$$

where $\boldsymbol{\varphi} = (\varphi_1, \ldots, \varphi_Q)$ are the set of replicated fields and $\hat{M}_{m,a}$ is the measurement operator acting on the $a$th replica. Next, we can use the Gaussian form of the measurements $\hat{M}_m$ to integrate them out obtaining the replicated action [2]

$$S_Q = \sum_{a=1}^Q S[\varphi_a] - \frac{\gamma}{2Q\pi^2} \int dx \sum_{b<c=1}^Q \cos(2(\varphi_b - \varphi_c)) + \ldots, \tag{9}$$

where we have neglected terms that are irrelevant in the sense of RG. With the normalization used in Eq. (1), the scaling dimension of $\cos(2\varphi)$ is $K$, which implies that the scaling dimension of $\cos(2\varphi_b - 2\varphi_c)$ is $2K$, leading to the RG equation $d\gamma/d\ell = (1-2K)\gamma$ at tree level. Thus, measurements are relevant for $K < 1/2$. When this is the case, the coupling $\gamma$ flows to the fixed point $\gamma \to \infty$ and measurements act to strongly lock the inter-replica fluctuations $\varphi_b - \varphi_c$ in the IR. Hereon, we will solely be interested in this strong-coupling limit of the above replica theory since it describes the IR physics of measurements.

The analogous case with a single replica is the boundary sine-Gordon perturbation of the type $-\lambda \int dx \cos(2\varphi)$. Here

$\lambda$ is relevant when $K < 1$ [1] and the perturbation pins the field to the minimum of the cosine potential $\varphi = p\pi$ with $p \in \mathbb{Z}$. However, for large but finite $\lambda$, phase-slip defects of integer multiples of $\pi$ arising due to the compact nature of the field are allowed; also sometimes referred to as instantons. The canonical realization of such a perturbation in condensed matter settings is that of impurity scattering in the Kane-Fisher setup [4] which was more recently adapted to the forced measurement of local density as discussed in Ref [2]. We shall say more about this when discussing phase correlators in Sec. III A.
The next step to deal with the replicated action (9) is to rotate into a basis,

$$\tilde{\boldsymbol{\varphi}} = U_Q \boldsymbol{\varphi}, \qquad (U_Q)_{ij} = \begin{cases} \delta_{ij} - \dfrac{1}{\sqrt{Q}(\sqrt{Q}-1)}, & i,j = 1,\dots,Q-1, \\ \dfrac{1}{\sqrt{Q}}, & i = Q \text{ or } j = Q. \end{cases} \tag{10}$$

that isolates the center of mass (c.o.m) degree of freedom from the relative ones, since the former is unaffected by measurements. Such a rotation trick was first used by Fradkin and Moore in the context of 2D quantum critical points [5] and later adapted in future works on the subject [6–9]. The matrix $U_Q$ can be constructed as a reflection that takes the vector $(1,\dots,1)_Q$ to the vector $(0,0,\dots,1)$ up to an overall constant. In this rotated basis, the above action takes the form

$$S_Q = \sum_{a=1}^{Q} S[\tilde{\varphi}_a] - \frac{\gamma}{2Q\pi^2}\int dx \left[\sum_{b<c<Q} \cos(2(\tilde{\varphi}_b - \tilde{\varphi}_c)) + \sum_{b=1}^{Q-1} \cos\left(2\tilde{\varphi}_b - \frac{2}{\sqrt{Q}-1}\sum_{c=1}^{Q-1}\tilde{\varphi}_c\right)\right]. \tag{11}$$

The upshot of this transformation is now clear: only $Q-1$ relative replicas enter the potential while the $Q$th c.o.m replica $\tilde{\varphi}_Q = Q^{-1/2}\sum_{i=1}^{Q}\varphi_i$ is left to freely fluctuate at the boundary. Equivalently, $\tilde{\varphi}_Q$ admits Neumann boundary conditions (NBCs). Consequently, we can now write the $\gamma \gg 1$ saddle point solution in this new basis as

$$\tilde{\boldsymbol{\varphi}} = \pi U_{Q-1}\boldsymbol{w}, \tag{12}$$

where $U_{Q-1}$ is the top-left $(Q-1)\times(Q-1)$ block of $U_Q$ and $\boldsymbol{w} \in \mathbb{Z}_{Q-1}$ enter due to the compactness of the fields. The above saddle is a replica version of the sine-Gordon saddle point mentioned earlier in this section. In a similar way to sine-Gordon, allowing $\gamma$ to be large but finite such that it is close to (but not at) the fixed point $\gamma \to \infty$, the theory can admit instanton or phase-slip solutions, now in replica space, as we show below.

### B. Replica Coulomb Gas

In what follows, it will be convenient to focus on the boundary theory since the bulk is unaffected by the measurements. Accordingly, integrating out the bulk, one obtains

$$S_{\text{bdry},Q} = \sum_{a=1}^{Q}\frac{1}{\pi K}\int\frac{dq}{2\pi}|q||\tilde{\varphi}_a(q)|^2 - \frac{\gamma}{2Q\pi^2}\int dx\, V(\tilde{\boldsymbol{\varphi}}), \tag{13}$$

where $\tilde{\varphi}_a(q)$ is the Fourier transform of the field $\tilde{\varphi}_a(x)$ and we use $V(\tilde{\boldsymbol{\varphi}})$ to denote the replica potential from Eq. (11). For large but finite coupling $\gamma$, one expects that the field configurations to be restricted to their minima (12). However, phase slip events or instantons that are tunnelings between different minima of $\tilde{\boldsymbol{\varphi}}$ are allowed, changing $\boldsymbol{w} \to \boldsymbol{w} + \Delta\boldsymbol{w}$. A controlled description of such instantons requires us to introduce a short-distance UV regularization. This regulator fixes the nonuniversal structure of the instanton core, while leaving the universal long-distance physics unchanged. A canonical choice is of the type $\frac{\lambda}{2}\int dx \sum_{a=1}^{Q}(\partial_x\tilde{\varphi}_a)^2$. With this UV regulator, we make an estimate of the action and the underlying length scale for the core of the instanton. A phase slip $\boldsymbol{w} \to \boldsymbol{w} + \Delta\boldsymbol{w}$ over a length scale $\ell$ costs $S_{\text{core}} \sim \frac{\lambda}{2\ell}\Delta\boldsymbol{w}^T T_{Q-1}\Delta\boldsymbol{w} + \gamma\ell$, where we have used $\Delta\tilde{\boldsymbol{\varphi}} \sim \Delta\boldsymbol{w}T_{Q-1}\Delta\boldsymbol{w}$ with $T_{Q-1} = U_{Q-1}^T U_{Q-1}$. Minimizing the core action w.r.t $\ell$ we obtain $\ell \sim \left(\frac{\lambda}{2\gamma}\Delta\boldsymbol{w}T_{Q-1}\Delta\boldsymbol{w}\right)^{1/2}$ which scales as $\ell \sim 1/\sqrt{\gamma}$ for $\gamma \gg 1$. The core action at this extremum scales as $S_{\text{core}} \sim \sqrt{\lambda\gamma\Delta\boldsymbol{w}T_{Q-1}\Delta\boldsymbol{w}}$ and so the fugacity $g$, i.e. the Boltzmann weight associated with the creation of an instanton core, scales as $g \sim e^{-\sqrt{\gamma}} \ll 1$ for large $\gamma$. Therefore, for $\gamma \gg 1$, the width of the instantons scales as $\ell \ll g^{-1}$, the typical separation between them, implying that the domain walls are rare and dilute in this limit. Accordingly, for $\gamma \gg 1$, we can approximate instanton cores as sharp jumps and use the ansatz

$$\tilde{\boldsymbol{\varphi}}(x) = \pi\sum_j U_{Q-1}\Delta\boldsymbol{w}_j\,\theta(x - x_j), \tag{14}$$

where $x_j$ denotes the position of the $j$-th instanton. Substituting this solution into the non-local action Eq. (13) one obtains

$$S_{\text{bdry},Q} = -\frac{2}{K}\sum_{j<k} \Delta\boldsymbol{w}_j T_{Q-1} \Delta\boldsymbol{w}_k \log|x_j - x_k|, \tag{15}$$

analogous to the Coulomb gas in the single replica case of the boundary sine-Gordon potential. The corresponding partition function takes the form

$$Z_{\text{slip}}(Q) = \sum_{n=0}^{\infty} \int_{x_1<\cdots<x_{2n}} \prod_{i=1}^{2n} dx_j \sum_{\Delta\boldsymbol{w}_j | \sum_j \Delta\boldsymbol{w}_j = 0} g(\Delta\boldsymbol{w}_i) \exp\left[\frac{2}{K}\sum_{j<k} \Delta\boldsymbol{w}_j^T T_{Q-1} \Delta\boldsymbol{w}_k \log|x_j - x_k|\right], \tag{16}$$

where $g(\Delta\boldsymbol{w}_i) \ll 1$ is the fugacity of instantons with a phase slip of $\Delta\boldsymbol{w}_i$, which provides a controlled expansion in the instanton density. We also consider periodic boundary conditions in writing the above which restricts $\sum_j \Delta\boldsymbol{w}_j = 0$ assuming we work in the zero magnetization sector for simplicity.

## II. DENSITY CORRELATOR

### A. Setup

In this section we present a detailed calculation of the moments of the density correlators,

$$\mathcal{N}_k(x) = \sum_m p_m [\langle \hat{n}(x)\hat{n}(0)\rangle_m - \langle \hat{n}(x)\rangle_m \langle \hat{n}(0)\rangle_m]^k, \tag{17}$$

using the replicated Coulomb-gas (16) derived in the previous section. To proceed, it is convenient to express the above correlator in the replica-field theory formalism as

$$\mathcal{N}_k(x) = \lim_{Q\to 1} \frac{1}{\pi^{2k} 2^k} \left\langle \prod_{i=1}^{k} (\partial\varphi_{2i-1}(x) - \partial\varphi_{2i}(x))(\partial\varphi_{2i-1}(0) - \partial\varphi_{2i}(0)) \right\rangle_{Z_{\text{slip}}(Q)}, \tag{18}$$

with $Q > 2k$, where we have restricted to the smooth part of density and have implicitly set the normalization $\lim_{Q\to 1} \sum_m p_m^Q = 1$. The above follows from re-writing the replica formulation of the mean (see Eq. (8)) as

$$\partial\varphi_1(x)\partial\varphi_1(0) - \partial\varphi_1(x)\partial\varphi_2(0) = 2^{-1}(\partial\varphi_1(x) - \partial\varphi_2(x))(\partial\varphi_1(0) - \partial\varphi_2(0)) \tag{19}$$

and raising it to the $k$th power, with each power in distinct pairs of replicas $(2i-1, 2i)$. Since we are now working in the rotated basis (10), we must re-write this operator in the same basis, which we do as follows:

$$[\partial\varphi_{2i-1}(x) - \partial\varphi_{2i}(x)][\partial\varphi_{2i-1}(0) - \partial\varphi_{2i}(0)] = \sum_{j,k=1}^{Q} [(U_Q)_{j,2i-1} - (U_Q)_{j,2i}][(U_Q)_{k,2i-1} - (U_Q)_{k,2i}]\partial\tilde{\varphi}_j(x)\partial\tilde{\varphi}_k(0) \tag{20}$$

$$= [(U_{Q-1}(\boldsymbol{e}_{2i-1} - \boldsymbol{e}_{2i})) \cdot \partial\tilde{\boldsymbol{\varphi}}(x)][(U_{Q-1}(\boldsymbol{e}_{2i-1} - \boldsymbol{e}_{2i})) \cdot \partial\tilde{\boldsymbol{\varphi}}(0)] \tag{21}$$

where in the last line we replace $U_Q \to U_{Q-1}$. This follows from having taken $(2i-1, 2i) < Q \ \ \forall\ i$ and that for $j, k = Q$, $(U_Q)_{Q,2i-1} - (U_Q)_{Q,2i} = 0 \ \ \forall\ i$, that is, inter-replica differences live purely in the relative sector which involves only the first $Q-1$ indices. Therefore, the expectation value to be evaluated in the rotated basis is

$$\mathcal{N}_k(x) = \frac{1}{\pi^{2k} 2^k} \left\langle \prod_{i=1}^{k} [(U_{Q-1}(\boldsymbol{e}_{2i-1} - \boldsymbol{e}_{2i})) \cdot \partial\tilde{\boldsymbol{\varphi}}(x)][(U_{Q-1}(\boldsymbol{e}_{2i-1} - \boldsymbol{e}_{2i})) \cdot \partial\tilde{\boldsymbol{\varphi}}(0)] \right\rangle_{Z_{\text{slip}}(Q)}. \tag{22}$$

Next, since $Z_{\text{slip}}$ in Eq. (16) is a perturbative expansion in $g(\Delta\boldsymbol{w})$, the first contribution to analyse is at $\mathcal{O}(g^0)$ where there are no instantons and we are *at* the $\gamma \to \infty$ fixed point with uniform DBC for $\tilde{\boldsymbol{\varphi}}$. At this order, density fluctuations arise solely from small oscillations of the replicated fields about the minima in Eq. (14). The replicas therefore decouple, and the replicated correlator in Eq. (18) reduces to

$$\mathcal{N}_k(x)|_{g^0} = \lim_{Q\to 1} \frac{1}{\pi^{2k}} \langle \partial\varphi(x)\partial\varphi(0)\rangle^k \sim x^{-4k}. \tag{23}$$

Here we have used the single-replica correlator for fluctuations about the pinned configuration, which decays as $x^{-4}$, as shown in Ref. [2]. We revisit this result in Sec. IV, where we provide a more formal argument using boundary conformal field theory (BCFT). We will see that the above scaling is often a sub-leading universal contribution to the correlator in comparison to the contribution at the sub-leading order in $g$, $\mathcal{O}(g^2)$. At this order, periodic boundary conditions and zero magnetization imply that one has a pair of phase slips of charges $\{+\Delta\boldsymbol{w}, -\Delta\boldsymbol{w}\}$ located at $x_1$ and $x_2$. Using Eq. (16), the partition function within this sector then reduces to

$$Z_{2-\text{slip}} = \sum_{\Delta\boldsymbol{w}\in\mathbb{Z}^{Q-1}} \exp\left(-2\chi \Delta\boldsymbol{w} T_{Q-1}\Delta\boldsymbol{w}\right), \tag{24}$$

where we have introduced $\chi = K^{-1}\log|x_1 - x_2| + \kappa(\gamma)$. In this 2-slip sector, we have picked the functional form of the fugacity to be $g(\Delta\boldsymbol{w}) = \exp\left[-\kappa(\gamma)\Delta\boldsymbol{w}T_{Q-1}\Delta\boldsymbol{w}\right]$, where $\kappa(\gamma)$ is some function of $\gamma$. This choice should be viewed as a particular UV regularization of the instanton core. Since the detailed functional form of the fugacity is set by short-distance physics, it can affect nonuniversal amplitudes and finite-distance corrections, but not the universal long-distance scaling that we discuss here. In addition, we expect higher order terms at $\mathcal{O}(g^{2n})$ that include $2n$ slips with $n > 1$ to produce universal contributions to the correlator that are at most as slow as the one produced at $\mathcal{O}(g^2)$. To see this, consider the $2n$ neutral phase slips in two possible configurations. If they are all placed at microscopic distances from 0 or $x$ where we evaluate the density correlator, those slips act as a single effective pair of phase slips of charges $\pm\Delta\boldsymbol{w}' \in \mathbb{Z}^{Q-1}$ at 0 and $x$ respectively, where $\Delta\boldsymbol{w}'$ is the sum of the constituent charges. But since we sum over all $\Delta\boldsymbol{w} \in \mathbb{Z}^{Q-1}$ at $\mathcal{O}(g^2)$ already, this does not introduce new contributions to the correlator. Instead if we place the phase slips away from $0, x$, say at some $x_i = u_i x$ with $u_i \sim \mathcal{O}(1)$, they incur a cost that introduces additional powers of $x$ and give sub-leading corrections to the density correlator.

Using Eq. (14) in this 2-slip sector, we get

$$\tilde{\boldsymbol{\varphi}}(y) = \pi U_{Q-1}\Delta\boldsymbol{w}[\theta(y - x_1) - \theta(y - x_2)], \tag{25}$$

and so we must evaluate the correlator

$$\mathcal{N}_k(x) \simeq \lim_{Q\to 1}\frac{1}{2^k}\left\langle \prod_{i=1}^{k}[(U_{Q-1}(\boldsymbol{e}_{2i-1} - \boldsymbol{e}_{2i}))\cdot(U_{Q-1}\Delta\boldsymbol{w})][(U_{Q-1}(\boldsymbol{e}_{2i-1} - \boldsymbol{e}_{2i}))\cdot(U_{Q-1}\Delta\boldsymbol{w})]\right\rangle_{Z_{2-\text{slip}}(Q)} \tag{26}$$

$$= \lim_{Q\to 1}\frac{1}{2^k}\left\langle \prod_{i=1}^{k}((\boldsymbol{e}_{2i-1} - \boldsymbol{e}_{2i})T_{Q-1}\Delta\boldsymbol{w})^2\right\rangle_{Z_{2-\text{slip}}(Q)} \tag{27}$$

$$= \lim_{Q\to 1}\frac{1}{2^k}\sum_{\Delta\boldsymbol{w}\in\mathbb{Z}^{Q-1}}\prod_{i=1}^{k}((\boldsymbol{e}_{2i-1} - \boldsymbol{e}_{2i})T_{Q-1}\Delta\boldsymbol{w})^2 e^{-2\chi\Delta\boldsymbol{w}T_{Q-1}\Delta\boldsymbol{w}} \tag{28}$$

$$= \lim_{Q\to 1}\frac{1}{2^k}\sum_{\Delta\boldsymbol{w}\in\mathbb{Z}^{Q-1}}\partial^2_{J_1}\dots\partial^2_{J_k} e^{-2\chi\Delta\boldsymbol{w}T_{Q-1}\Delta\boldsymbol{w} + J_1(\boldsymbol{e}_1-\boldsymbol{e}_2)T_{Q-1}\Delta\boldsymbol{w}+\dots J_k(\boldsymbol{e}_{2k-1}-\boldsymbol{e}_{2k})T_{Q-1}\Delta\boldsymbol{w}}\Big|_{J_1,\dots,J_k=0} \tag{29}$$

where we have dropped overall non-universal factors, including signs. $(\boldsymbol{e}_i)_a = \delta_{ai}$ is a unit vector and only $x_1 = 0$ and $x_2 = x > 0$ contribute due to the step functions in $\tilde{\boldsymbol{\varphi}}(y)$. In the last step, we have introduced the sources $J_i$, which allow us to perform the analytic continuation described below.

### B. Analytic Continuation

In order to take the replica limit $Q \to 1$, we must analytically continue the above expression. For brevity, we define a variable $\Lambda(Q)$ that includes only the relevant pieces of (29) for analytic continuation, with

$$\Lambda(Q) = \sum_{\Delta\boldsymbol{w}\in\mathbb{Z}^{Q-1}} e^{-2\chi\Delta\boldsymbol{w}T_{Q-1}\Delta\boldsymbol{w} + J_1(\boldsymbol{e}_1-\boldsymbol{e}_2)T_{Q-1}\Delta\boldsymbol{w}+\dots J_k(\boldsymbol{e}_{2k-1}-\boldsymbol{e}_{2k})T_{Q-1}\Delta\boldsymbol{w}}. \tag{30}$$

We start by completing the square,

$$-2\chi\Delta\boldsymbol{w}T_{Q-1}\Delta\boldsymbol{w} + \Gamma T_{Q-1}\Delta\boldsymbol{w} = -2\chi\left(\Delta\boldsymbol{w} - \frac{\Gamma}{4\chi}\right)T_{Q-1}\left(\Delta\boldsymbol{w} - \frac{\Gamma}{4\chi}\right) + \frac{1}{8\chi}\Gamma^T T_{Q-1}\Gamma, \tag{31}$$

where $\sum_{i=1}^{k} J_i(\boldsymbol{e}_{2i-1}-\boldsymbol{e}_{2i})=\Gamma$. Using $\Gamma^T T_{Q-1}\Gamma = 2\sum_{i=1}^{k} J_i^2$, we can then write,

$$\Lambda(Q) = \sum_{\Delta\boldsymbol{w}\in\mathbb{Z}^{Q-1}} e^{-2\chi\Delta\boldsymbol{w}T_{Q-1}\Delta\boldsymbol{w}+J_1(\boldsymbol{e}_1-\boldsymbol{e}_2)T_{Q-1}\Delta\boldsymbol{w}+\ldots J_k(\boldsymbol{e}_{2k-1}-\boldsymbol{e}_{2k})T_{Q-1}\Delta\boldsymbol{w}}$$
$$= \sum_{\Delta\boldsymbol{w}\in\mathbb{Z}^{Q-1}} e^{\frac{1}{4\chi}\sum_{i=1}^{k}J_i^2} e^{-2\chi\left(\Delta\boldsymbol{w}-\frac{\Gamma}{4\chi}\right)T_{Q-1}\left(\Delta\boldsymbol{w}-\frac{\Gamma}{4\chi}\right)}. \quad (32)$$

Re-casting $\Lambda(Q)$ in this way now allows us to apply the multi-dimensional reciprocal formula [10] and analytically continue as follows [9]:

$$\Lambda(Q) = \sqrt{Q}\left(\frac{\pi}{2\chi}\right)^{\frac{Q-1}{2}} \sum_{\Delta\boldsymbol{w}\in\mathbb{Z}^{Q-1}} e^{\frac{1}{4\chi}\sum_{i=1}^{k}J_i^2} e^{-\frac{\pi^2}{2\chi}\Delta\boldsymbol{w}T_{Q-1}^{-1}\Delta\boldsymbol{w}-\frac{i\pi}{2\chi}\Delta\boldsymbol{w}^T\Gamma},$$
$$= \sqrt{Q}\left(\frac{\pi}{2\chi}\right)^{\frac{Q-1}{2}} \sum_{\Delta\boldsymbol{w}\in\mathbb{Z}^{Q-1}} e^{\frac{1}{4\chi}\sum_{i=1}^{k}J_i^2} e^{-\frac{\pi^2}{2\chi}\left(\sum_{a=1}^{Q-1}w_a^2\right)-\frac{\pi^2}{2\chi}\left(\sum_{a=1}^{Q-1}w_a\right)^2-\frac{i\pi}{2\chi}\Delta\boldsymbol{w}^T\Gamma},$$
$$= \sqrt{Q}\left(\frac{\pi}{2\chi}\right)^{\frac{Q-1}{2}} e^{\frac{1}{4\chi}\sum_{i=1}^{k}J_i^2} \int_0^{\pi}\frac{d\delta}{\pi}\sum_{l\in\mathbb{Z}} e^{-\frac{\pi^2}{2\chi}l^2+2il\delta} \sum_{\Delta\boldsymbol{w}\in\mathbb{Z}^{Q-1}} e^{-\frac{\pi^2}{2\chi}\left(\sum_{a=1}^{Q-1}w_a^2\right)+2i\delta\sum_a w_a-\frac{i\pi}{2\chi}\sum_a\Gamma_a w_a},$$
$$= \sqrt{\frac{2Q\chi}{\pi}}\int_0^{\pi}\frac{d\delta}{\pi}\,\mathcal{Z}_\chi(\delta)^{Q-2k}\prod_{i=1}^{k}\mathcal{Z}_\chi(\delta,J_i)\,\mathcal{Z}_\chi(\delta,-J_i)\,, \quad (33)$$

where we have defined

$$\mathcal{Z}_\chi(\delta,J) = \sum_{w\in\mathbb{Z}} e^{-2\chi\left(w+\frac{\delta}{\pi}\right)^2+J\left(w+\frac{\delta}{\pi}\right)}, \quad (34)$$

with $\mathcal{Z}_\chi(\delta,0)\equiv\mathcal{Z}_\chi(\delta)$. The above steps use the fact that $\Delta\boldsymbol{w}T_{Q-1}^{-1}\Delta\boldsymbol{w} = \sum_{a=1}^{Q-1}\Delta w_a^2 + \left(\sum_{a=1}^{Q-1}\Delta w_a\right)^2$ and the identity

$$\sum_{w\in\mathbb{Z}}\exp\left[-\pi a w^2+2\pi i b w\right] = \frac{1}{\sqrt{a}}\sum_{w\in\mathbb{Z}}\exp\left[-\frac{\pi}{a}(w+b)^2\right]. \quad (35)$$

The analytically continued Eq. (33) is now valid for all $Q$ and $k$, including for $Q<2k$. We can now take the replica limit $Q\to 1$ in Eq. (29) and write $\mathcal{N}_k(x)$ as

$$\mathcal{N}_k(x) = \sqrt{\frac{2\chi}{\pi}}\int_0^{\pi}\frac{d\delta}{\pi}e^{-F_\chi(\delta)}\left[-\partial_J^2 F_\chi(\delta,J)|_{J=0}\right]^k, \quad (36)$$

where the $k$-derivatives in Eq. (29) factorize, and we have defined $F_\chi(\delta,J) = -\log\mathcal{Z}_\chi(\delta,J)$ as a free energy associated to the partition function $\mathcal{Z}_\chi(\delta,J)$. The above expression is Eq. 5 in the main text where we have only kept the universal part in the main text by replacing $\chi = K^{-1}\log x+\kappa(\mu)\to K^{-1}\log x, \mathcal{Z}_\chi\to\mathcal{Z}_x$. Here the integral

$$\sqrt{\frac{2\chi}{\pi}}\int_0^{\pi}\frac{d\delta}{\pi}\,e^{-F_\chi(\delta)} = 1, \quad (37)$$

and therefore the factor $\mathcal{W}^{-1}e^{-F_\chi(\delta)}$ with $\mathcal{W}=\pi/\sqrt{2\chi}$ is a normalized probability distribution that acts as the Born distribution in the IR, with $\delta\in[0,\pi)$ parametrizing the coarse-grained measurement outcomes. Upon taking the derivatives w.r.t $J$ and then setting $J=0$, we can rewrite Eq. (36) as

$$\mathcal{N}_k(x) = \sqrt{\frac{2\chi}{\pi}}\int_0^{\pi}\frac{d\delta}{\pi}\,\mathcal{Z}_\chi(\delta)\left[\frac{\sum_{v\in\mathbb{Z}}\left(v+\frac{\delta}{\pi}\right)^2\exp\left[-2\chi\left(v+\frac{\delta}{\pi}\right)^2\right]}{\mathcal{Z}_\chi(\delta)} - \left(\frac{\sum_{v\in\mathbb{Z}}\left(v+\frac{\delta}{\pi}\right)\exp\left[-2\chi\left(v+\frac{\delta}{\pi}\right)^2\right]}{\mathcal{Z}_\chi(\delta)}\right)^2\right]^k. \quad (38)$$

Therefore, we can interpret $\mathcal{N}_k(x)$ as being the $k$-th moment of variance of charge $(v+\delta/\pi)$ w.r.t partition function $\mathcal{Z}_\chi(\delta)$ for the outcome $\delta$, averaged over the normalized effective Born-weight (see Eq. (37)). The appearance of variance of the charge here is natural since in the two-slip sector, the smooth density $\sim\partial_x\varphi$ measures the charge of the phase-slip (see from Eq. (25) for example). We discuss this further in Sec. IV in the context of a more general boundary conformal field theory (BCFT) interpretation.

### C. Asymptotics and Scaling

We now extract the asymptotic large $\chi$, or equivalently, large $x$ scaling of (38). Let us start by introducing the variable $u = \delta/\pi$ and write

$$\mathcal{N}_k(x) = \sqrt{\frac{2\chi}{\pi}} 2 \int_0^{1/2} du \, \mathcal{Z}_\chi(\pi u) \left[ \frac{\sum_{v\in\mathbb{Z}}(v+u)^2 e^{-2\chi(v+u)^2}}{\mathcal{Z}_\chi(\pi u)} - \left( \frac{\sum_{v\in\mathbb{Z}}(v+u) e^{-2\chi(v+u)^2}}{\mathcal{Z}_\chi(\pi u)} \right)^2 \right]^k, \tag{39}$$

where we can only focus on $u \in [0, 1/2]$ since Eq. (38) is symmetric about $\delta = \pi/2$ [this comes from properties such as $\mathcal{Z}_\chi(\pi u)\mathcal{Z}_\chi(\pi(1-u))$]. Next, since $e^{-\chi} \ll 1$ for $x \gg 1$, we can only focus on the first few leading terms; these are those with $v = -1, 0$. This reduces the above expression to

$$\mathcal{N}_k(x) \simeq \sqrt{\frac{8\chi}{\pi}} \int_0^{1/2} du \, e^{-2\chi u^2} e^{-2k\chi(1-2u)}. \tag{40}$$

We proceed with a saddle point analysis of the simplified expression above. Analytically continuing to $k \in \mathbb{R}$, the saddle for $0 < k < 1/2$ is when $u_* = k$. This returns

$$\mathcal{N}_k(x) \simeq \sqrt{\frac{8\chi}{\pi}} e^{-2\chi k(1-k)} \int_0^{1/2} du \, e^{-2\chi(u-k)^2} \sim e^{-2\chi k(1-k)} \sim x^{-\frac{2}{K}k(1-k)} \quad , \quad k < 1/2. \tag{41}$$

For $k < 1/2$, the saddle $u_* = k$ lies inside $u < 1/2$. Near $u \simeq 1/2$ however, the $v = -1, 0$ sectors in $\mathcal{Z}_\chi(\pi u)$ become degenerate and the above analysis changes. Taking $u = \frac{1}{2} - \frac{s}{\chi}$, Eq. (39) becomes

$$\mathcal{N}_k(x) \simeq \sqrt{\frac{8}{\pi}} \frac{e^{-\chi/2}}{\sqrt{\chi}} \int_0^{\chi/2} ds \, \frac{e^{-2s^2/\chi}}{(e^{2s} + e^{-2s})^{2k-1}}. \tag{42}$$

For $k > 1/2$ above integral is dominated by $s \sim O(1)$ for which $e^{-2s^2/\chi} \simeq 1$ which leads to the whole integral being $\mathcal{O}(1)$ in $\chi$. This results in a $k-$independent scaling,

$$\mathcal{N}_k(x) \sim \frac{x^{-\frac{1}{2K}}}{\sqrt{\log x}} \quad , \quad k > 1/2, \tag{43}$$

whereas, for $k = 1/2$, the integral scales as $\mathcal{O}(\sqrt{\chi})$, leading to $\mathcal{N}_k(x)|_{k=1/2} \sim x^{-\frac{1}{2K}}$. Putting this all together (including the contribution at $\mathcal{O}(g^0)$ in Eq. (23)), we have

$$\mathcal{N}_k(x) \sim \begin{cases} x^{-\frac{2k(1-k)}{K}} + x^{-4k}, & 0 < k \leq 1/2 \\ \frac{x^{-\frac{1}{2K}}}{\sqrt{\log x}} + x^{-4k}, & k > 1/2 \end{cases}. \tag{44}$$

## III. PHASE CORRELATORS

### A. Phase-Slips as Vertex Operators

In this section we present a detailed calculation of the moments of the phase correlators,

$$\Theta_k(x) = \sum_m p_m |\langle e^{i[\theta(x)-\theta(0)]} \rangle_m|^{2k}. \tag{45}$$

We begin by briefly re-formulating the *forced* measurement scenario in Ref. [2], where the local density was measured, imprinting a CDW onto the ground state of the TLL. The relevant boundary theory with the bulk integrated out is [2],

$$S_{\text{bdry}} = \frac{1}{\pi K} \int \frac{dq}{2\pi} \, |q| |\tilde{\varphi}(q)|^2 - \lambda \int dx \, \cos 2\varphi, \tag{46}$$

where $\tilde{\varphi}(q)$ is the Fourier transform $\tilde{\varphi}(q) = \int dx e^{-iqx}\varphi(x)$ of the field $\varphi$, and $\lambda$ is the measurement strength. When $K < 1$, the above perturbation is relevant and, in the IR, pins the field $\varphi = w\pi$, where $w \in \mathbb{Z}$. The above is the well-known boundary sine-Gordon theory. Consequently, Ref. [2] shows that in this boundary theory

$$\langle e^{i[\theta(0)-\theta(x)]}\rangle \sim x^{-1/K}, \tag{47}$$

in the IR, at the $\lambda \to \infty$ fixed point. This differs from the corresponding correlator in the unmeasured TLL, $\langle e^{i[\theta(x)-\theta(0)]}\rangle_0 \sim x^{-1/(2K)}$ , showing that the measurement alters its long-distance scaling. Here we treat this result as the fundamental correlator at the boundary, with the more general version of it being

$$\langle e^{iw[\theta(0)-\theta(x)]}\rangle \sim x^{-w^2/K}, \tag{48}$$

where $w \in \mathbb{Z}$. Physically, one should think of the vertex operator $e^{\mp iw\theta(x)}$ as acting to induce phase slips of $\pm w\pi$ in the field $\varphi$.

In the replicated theory Eq. (13), we have $Q$ independent copies of the boundary sine-Gordon theory with DBC on $\tilde{\boldsymbol{\varphi}}$ and NBC on $\tilde{\varphi}_Q$. The main constraint for the fields $\tilde{\boldsymbol{\varphi}}$ however is that the phase slips are not arbitrary integer multiples of the individual period $\pi$, namely, $\pi\boldsymbol{q} = (q_1, \ldots, q_{Q-1}) \in \mathbb{Z}^{Q-1}$, but are constrained to be of the form $\pi U_{Q-1}\Delta\boldsymbol{w}$ with $\Delta\boldsymbol{w} \in \mathbb{Z}^{Q-1}$. Therefore, in the replicated theory, the allowed replica vertex operators are of the type $e^{i(U_{Q-1}\Delta\boldsymbol{w})\cdot\tilde{\boldsymbol{\theta}}}$, with the two-point function

$$\left\langle e^{i(U_{Q-1}\Delta\boldsymbol{w})\cdot\tilde{\boldsymbol{\theta}}(x)} e^{-i(U_{Q-1}\Delta\boldsymbol{w})\cdot\tilde{\boldsymbol{\theta}}(0)}\right\rangle_0 \sim x^{-\frac{1}{K}\Delta\boldsymbol{w}T_{Q-1}\Delta\boldsymbol{w}}, \tag{49}$$

where $T_{Q-1} = U_{Q-1}^T U_{Q-1}$, $\langle\ldots\rangle_0$ denotes an expectation value w.r.t the replicated boundary sine-Gordon theory (13) with DBC (12) on $\tilde{\boldsymbol{\varphi}}$, and where we ignore sub-leading terms. The above follows from a simple generalization of (48) where we have used the dot product $(U_{Q-1}\Delta\boldsymbol{w}).(U_{Q-1}\Delta\boldsymbol{w})$. This correlator then suggests a re-writing of the replicated coulomb gas partition function (16) as

$$Z_{\rm slip} \simeq \sum_n \int_{x_1<\cdots<x_{2n}} \prod_{j=1}^{2n} dx_j \sum_{\{\Delta\boldsymbol{w}_j\}:\,\sum_\Delta \boldsymbol{w}_j=0} \prod_{j=1}^{2n} g(\Delta\boldsymbol{w}_j) \left\langle \prod_{j=1}^{2n} e^{i(U_{Q-1}\Delta\boldsymbol{w}_j)\cdot\tilde{\boldsymbol{\theta}}_+(x_j)}\right\rangle_0 \left\langle \prod_{j=1}^{2n} e^{i(U_{Q-1}\Delta\boldsymbol{w}_j)\cdot\tilde{\boldsymbol{\theta}}_-(x_j)}\right\rangle_0, \tag{50}$$

where we have two copies $\tilde{\boldsymbol{\theta}}_\pm$ associated to the evolution of the bra (+) and the ket (−). The above returns Eq. (16) upon using Eq. (49) and applying Wick's theorem. Each branch contributes a power of $x$ proportional to $1/K$, and their product gives the coefficient $2/K$ appearing in the replicated Coulomb gas (see Eq. (15)). Such a re-writing is natural since the Coulomb gas (16) is indeed a dilute gas of phase slips that are implemented by vertex operators of the dual field. For the single-replica case, an analogous re-writing is the standard strong-coupling representation of the boundary sine-Gordon theory (46) [1]: once the field $\varphi$ is pinned, the approach to the Dirichlet IR fixed point is described by dual-field vertex operators that create tunnelling events between adjacent minima. Eq. (50) is then the replicated analog of this phase-slip or instanton expansion where we perturb the IR fixed point having DBC (12) with the allowed vertex operators of the replicated dual field $\tilde{\boldsymbol{\theta}}$.

### B. Calculation and Scaling

With these observations in mind, we can now evaluate the correlator (45). In the replica formalism, we can write

$$\Theta_k(x) = \lim_{Q\to 1}\left\langle \prod_{i=1}^k e^{i[\theta_{2i-1}(x)-\theta_{2i}(x)]} e^{-i[\theta_{2i-1}(0)-\theta_{2i}(0)]}\right\rangle_{Z_{\rm slip}}, \tag{51}$$

with $2k < Q$. In the rotated basis, it reads

$$\prod_{i=1}^k e^{i[\theta_{2i-1}(x)-\theta_{2i}(x))]} e^{-i[\theta_{2i-1}(0)-\theta_{2i}(0))]} = \exp\left[i\sum_{a=1}^k\left(\sum_{\mu=1}^{Q-1}(U_{\mu,2a-1}-U_{\mu,2a})\left[\tilde{\theta}_\mu(x)-\tilde{\theta}_\mu(0)\right]\right)\right] \tag{52}$$

$$= \exp\left[iU_{Q-1}\boldsymbol{q}_k\cdot(\tilde{\boldsymbol{\theta}}(x)-\tilde{\boldsymbol{\theta}}(0))\right], \tag{53}$$

with the replica charge $\boldsymbol{q}_k = \sum_{i=1}^k(\boldsymbol{e}_{2i-1}-\boldsymbol{e}_{2i})$, where $(\boldsymbol{e}_i)_a = \delta_{i,a}$ is a unit vector along $i$. Note that just like the density correlator in Sec. II A, the above correlator also lives entirely in the relative sector, that is, involving only

the first $Q-1$ relative indices in its replicated representation. While such a choice of correlator is not necessary, it is nonetheless useful to isolate universal effects of measurements since any correlator with a c.o.m component carries universal contributions that are not induced due to measurements and are therefore not of interest in this work.

Next, we evaluate (51) up to $\mathcal{O}(g^2)$, where the restriction to this order is justified below. At $\mathcal{O}(g^0)$, we are at the $\gamma \to \infty$ fixed point with DBC on $\tilde{\boldsymbol{\varphi}}$. The correlator at this order is given by:

$$\Theta_k(x)|_{g^0} = \lim_{Q\to 1} \left\langle \exp\left[iU_{Q-1}\boldsymbol{q}_k \cdot (\tilde{\boldsymbol{\theta}}(x) - \tilde{\boldsymbol{\theta}}(0))\right]\right\rangle_0 \sim x^{-\frac{1}{K}\boldsymbol{q}_k T_{Q-1}\boldsymbol{q}_k} = x^{-\frac{2k}{K}}, \tag{54}$$

where we used Eq. (49). In addition, $\boldsymbol{q}_k T_{Q-1}\boldsymbol{q}_k = 2k$ and so the replica trick is trivial, similar to the $\mathcal{O}(g^0)$ contribution for the density correlator we saw in Sec. II C. Indeed, at this order, there are no instanton fluctuations around the uniform Dirichlet boundary condition. The scaling exponent is therefore linear in $k$, corresponding to simple scaling rather than multifractal behavior. We next consider the leading instanton correction at order $\mathcal{O}(g^2)$.Using the 2-slip sector (see Eq. (24)) with phase-slips of charges $+\Delta\boldsymbol{w}$ and $-\Delta\boldsymbol{w}$, we can express the correlator as

$$\Theta_k(x)|_{g^2} = \lim_{Q\to 1} \sum_{\Delta\boldsymbol{w}\in\mathbb{Z}^{Q-1}} g(\Delta\boldsymbol{w})^2 \left\langle e^{i(U_{Q-1}\Delta\boldsymbol{w})\cdot\tilde{\boldsymbol{\theta}}_+(x)} e^{-i(U_{Q-1}\Delta\boldsymbol{w})\cdot\tilde{\boldsymbol{\theta}}_+(0)} e^{iU_{Q-1}\boldsymbol{q}_k\cdot(\tilde{\boldsymbol{\theta}}(x)-\tilde{\boldsymbol{\theta}}(0))}\right\rangle_0$$
$$\left\langle e^{i(U_{Q-1}\Delta\boldsymbol{w})\cdot\tilde{\boldsymbol{\theta}}_-(x)} e^{-i(U_{Q-1}\Delta\boldsymbol{w})\cdot\tilde{\boldsymbol{\theta}}_-(0)}\right\rangle_0 . \tag{55}$$

Here we have used the 2-slip version of the re-writing (50) and have placed the external replica operator (53) in the bra $(+)$ branch of the partition function. One could have equally placed it in the ket, and the result would not change. The two vertex operators inserted near $x$, $e^{i(U_{Q-1}\Delta\boldsymbol{w})\cdot\tilde{\boldsymbol{\theta}}_+(x)}$ and $e^{iU_{Q-1}\boldsymbol{q}_k\cdot\tilde{\boldsymbol{\theta}}(x)}$, should be understood through their short-distance operator product expansion (OPE). More generally, two nearby vertex operators $V_\alpha = e^{i\alpha\boldsymbol{q}}$ fuse into a vertex operator whose charge is the sum of the two, $V_\alpha(x)V_\beta(x+\epsilon) \underset{\epsilon\to 0}{\sim} V_{\alpha+\beta}(x)$, up to short-distance prefactors and descendants, with $a \sim \epsilon \ll L$, where $a$ is the ultraviolet cutoff. As a result, one can simplify the above correlator as

$$\Theta_k(x)|_{g^2} = \lim_{Q\to 1} \sum_{\Delta\boldsymbol{w}\in\mathbb{Z}^{Q-1}} g(\Delta\boldsymbol{w})^2 \left\langle e^{iU_{Q-1}(\Delta\boldsymbol{w}+\boldsymbol{q}_k)\cdot\tilde{\boldsymbol{\theta}}_+(x)} e^{-iU_{Q-1}(\Delta\boldsymbol{w}+\boldsymbol{q}_k)\cdot\tilde{\boldsymbol{\theta}}_+(0)}\right\rangle_0 \left\langle e^{i(U_{Q-1}\Delta\boldsymbol{w})\cdot\tilde{\boldsymbol{\theta}}_-(x)} e^{-i(U_{Q-1}\Delta\boldsymbol{w})\cdot\tilde{\boldsymbol{\theta}}_-(0)}\right\rangle_0 \tag{56}$$

$$= \lim_{Q\to 1} \sum_{\Delta\boldsymbol{w}\in\mathbb{Z}^{Q-1}} \exp\left[-\chi(\Delta\boldsymbol{w}+\boldsymbol{q}_k)T_{Q-1}(\Delta\boldsymbol{w}+\boldsymbol{q}_k) - \chi(\Delta\boldsymbol{w}T_{Q-1}\Delta\boldsymbol{w})\right]. \tag{57}$$

Here we remark that we have placed the two phase slips with charges $\{\Delta\boldsymbol{w}, -\Delta\boldsymbol{w}\}$ at 0 and $x$, where the replica two-point function is evaluated. The reason follows the same logic used above for the density correlator, and it explains both why this placement gives the slowest contribution at $\mathcal{O}(g^2)$ and why higher-order phase-slip configurations do not generate a slower universal decay. Placing the two slips macroscopically far from 0 and $x$ does not produce a slower universal decay of the phase correlator than placing them at 0 and $x$. When phase slips with charges $\{\Delta\boldsymbol{w}, -\Delta\boldsymbol{w}\}$ are inserted microscopically close to 0 and $x$, they fuse with the phase slips carrying charges $\{\boldsymbol{q}_k, -\boldsymbol{q}_k\}$, as shown in Eq. (56). The inserted phase slips can therefore screen these charges, producing an effective vertex operator with a potentially smaller scaling dimension than in the absence of phase slips. By contrast, if the additional phase slips are placed macroscopically far from 0 and $x$, say at $x_i = u_i x$ with $u_i \sim \mathcal{O}(1)$, no such screening occurs at long distances. Instead, the phase slips incur their own scaling cost, introducing additional powers of $x$. Their contributions are therefore expected to decay faster than both the fixed-point contribution and the contribution obtained when the phase slips are placed close to 0 and $x$. Higher-order phase-slip configurations likewise do not produce a slower decay of the replica correlator, for reasons similar to those discussed for the density correlator. If the $2n$ neutral phase slips lie within microscopic distances of 0 and $x$, the slips near each point fuse into a single effective phase slip. The resulting pair carries charges $\pm\Delta\boldsymbol{w}' \in \mathbb{Z}^{Q-1}$, where $\Delta\boldsymbol{w}'$ is the sum of the constituent charges near either point. Since the $\mathcal{O}(g^2)$ contribution already includes a sum over all $\Delta\boldsymbol{w} \in \mathbb{Z}^{Q-1}$, these higher-order clusters at 0 and $x$ do not generate any new universal scaling beyond that already present at $\mathcal{O}(g^2)$. Alternatively, any additional slips that remain macroscopically separated from 0 and $x$ incur further scaling costs and therefore give subleading corrections.

We must now analytically continue the above sum. We do it by first completing the square of the quadratic form involving $\Delta\boldsymbol{w}$ as

$$(\Delta\boldsymbol{w}+\boldsymbol{q}_k)T_{Q-1}(\Delta\boldsymbol{w}+\boldsymbol{q}_k) + \Delta\boldsymbol{w}T_{Q-1}\Delta\boldsymbol{w} = 2\left(\Delta\boldsymbol{w}+\frac{\boldsymbol{q}_k}{2}\right)T_{Q-1}\left(\Delta\boldsymbol{w}+\frac{\boldsymbol{q}_k}{2}\right) + \frac{1}{2}\boldsymbol{q}_k T_{Q-1}\boldsymbol{q}_k. \tag{58}$$

Using $\boldsymbol{q}_k T_{Q-1}\boldsymbol{q}_k = 2k$, we have

$$\Theta_k(x)|_{g^2} = \lim_{Q\to 1} e^{-k\chi} \sum_{\Delta\boldsymbol{w}\in\mathbb{Z}^{Q-1}} \exp\left[-2\chi\left(\Delta\boldsymbol{w}+\frac{\boldsymbol{q}_k}{2}\right)T_{Q-1}\left(\Delta\boldsymbol{w}+\frac{\boldsymbol{q}_k}{2}\right)\right]. \tag{59}$$

The analytic continuation proceeds as in Eqs. (32)-(33), after completing the square in Eq. (31). We directly give the resulting expression in the replica limit:

$$\Theta_k(x)|_{g^2} = \sqrt{\frac{2\chi}{\pi}} \int_0^\pi \frac{d\delta}{\pi} \, \mathcal{Z}_\chi(\delta) \left[ e^{-\frac{\chi}{2}} \frac{\mathcal{Z}_\chi(\delta+\pi/2)}{\mathcal{Z}_\chi(\delta)} \right]^{2k}, \tag{60}$$

where $\mathcal{Z}_\chi(\delta) \equiv \mathcal{Z}_\chi(\delta, 0)$ is given in Eq. (34). Extracting the leading universal decay from the above integral in the limit of large $\chi$, or equivalently large $x$, gives an asymptotic behavior similar to Sec. II C

$$\Theta_k(x)|_{g^2} \sim \begin{cases} x^{-\frac{2k}{K}(1-k)}, & 0 < k \leq 1/2, \\ \dfrac{x^{-\frac{1}{2K}}}{\sqrt{\log x}}, & k > 1/2. \end{cases} \tag{61}$$

Since the contribution at $\mathcal{O}(g^0)$, $\sim x^{-2k/K}$ decays faster than the above $\mathcal{O}(g^2)$ decay for all $k$, the above is indeed the leading universal decay of the phase correlator at all orders, since higher orders in $g$ lead to sub-leading corrections.

## IV. DETAILS ON THE GENERAL BCFT PICTURE

In this section we explicitly justify the general picture proposed in the main text in the context of the TLL, where the TLL is treated as the free-boson CFT. In particular, we show that the averaged correlation functions derived in the preceding sections can be obtained alternatively through an averaging prescription that requires two ingredients: the ensemble of conformally invariant boundary conditions (CIBCs) reached under the measurement-induced boundary RG flow, and the BCFT data associated with this ensemble.

We begin with the doubled-state representation of the ground-state density matrix $|\rho_0\rangle\rangle = |\psi_0\rangle\,|\psi_0\rangle^*$, where we refer to $|\psi_0\rangle$ ($|\psi_0\rangle^*$) as the ket (bra) copy. In the Euclidean path integral representation, these two copies can be viewed as lying on the two sides of the imaginary-time slice $\tau = 0$, namely $\tau = 0^-$ and $\tau = 0^+$. The measurement perturbation Eq. (3) acts along this slice and couples the two copies through their common measurement record. Folding the path integral along $\tau = 0$ therefore maps the measurement defect to a *boundary* perturbation of the doubled conformal field theory CFT. Next, we posit that the density measurements considered here (see Eq. (3)), when relevant, drive a boundary RG flow to the boundary fixed point of *uniform* Dirichlet boundary conditions (DBC) for the field $\varphi$ each of the $\tau = 0^\pm$ slices. The flow to Dirichlet boundary conditions for $\varphi$ is clear in the forced case [2] and more generally is consistent with symmetry, since density measurements preserve the global $U(1)$ symmetry $\theta \to \theta + \alpha$. This is also explained by the relevance and strong-coupling structure of the boundary perturbation.

The resulting free-boson BCFT is characterized by its boundary operator content, which is encoded in the (open-channel) annulus partition [11]

$$Z_{DD}(\delta) = \frac{1}{\eta(q)} \sum_{w\in\mathbb{Z}} q^{\frac{1}{2K}(w+\frac{\delta}{\pi})^2}, \tag{62}$$

where we consider an annulus of width 1 and length $L$ such that $q = e^{-\pi L}$. Here $\delta = \delta_1 - \delta_2$ is the difference between the DBC imposed along the two edges of the annulus that are unit distance apart. Each term in the above partition function encodes the scaling dimensions of the allowed boundary operators. The terms in the sum $\sum_{w\in\mathbb{Z}}$ encode the scaling dimensions $\Delta_{\delta,w} = (2K)^{-1}(w+\delta/\pi)^2$ of boundary condition changing (BCC) operators, which are identified with vertex operators $\hat{O}_{\delta,w}(x) = e^{i(w+\delta/\pi)\theta(x)}$ that change the DBC by $\delta + \pi w$ across $x$. In light of the discussion in the previous sections, these are precisely the phase-slip operators. The pre-factor $\eta(q)^{-1} = q^{-1/24}\sum_{n=0}^\infty p(n) q^n$ is a generating function for the partition number $p(n)$ and generates the oscillator descendants with scaling dimension $\Delta_{\delta,w} + n$. These generators are identified with the boundary operators $\prod_i \partial_x^{(m_i)}\theta$ where $\sum_i m_i = n$. The factor $q^{-1/24}$ corresponds to the ground-state (Casimir) energy.

As an example, consider the following correlator involving an arbitrary operator $\hat{A}$:

$$\mathcal{A}_k(x) = \sum_m p_m [\langle \hat{A}(x)\hat{A}(0)\rangle_m - \langle \hat{A}(x)\rangle_m \langle \hat{A}(0)\rangle_m]^k, \tag{63}$$

We claim that the universal decay of the above measurement-averaged correlator can be obtained by averaging over correlators evaluated with a DBC background with possible phase-slip BCCs $\hat{O}_{\delta,w}$ inserted along the boundary, with

weights corresponding to the correlator of the corresponding BCC insertions, appropriately normalized. Specifically, for the above measurement averaged two-point correlator,

$$\mathcal{A}_k(x) \sim \frac{\int_0^\pi d\delta\ \mathcal{Z}_x(\delta)\ [C_x(\delta, \hat{A})]^k}{\int_0^\pi d\delta\ \mathcal{Z}_x(\delta)}, \tag{64}$$

where,

$$C_x(\delta, \hat{A}) = \frac{1}{\mathcal{Z}_x(\delta)} \sum_{w\in\mathbb{Z}} \langle \hat{\mathcal{O}}_{\delta,w}(0)\hat{A}(0)\hat{A}(x)\hat{\mathcal{O}}_{\delta,-w}(x)\rangle_0 - \frac{1}{\mathcal{Z}_x^2(\delta)} \left(\sum_{u\in\mathbb{Z}} \langle \hat{\mathcal{O}}_{\delta,u}(0)\hat{A}(x)\hat{\mathcal{O}}_{\delta,-u}(x)\rangle_0\right) \left(\sum_{v\in\mathbb{Z}} \langle \hat{\mathcal{O}}_{\delta,v}(0)\hat{A}(0)\hat{\mathcal{O}}_{\delta,-v}(x)\rangle_0\right) \tag{65}$$

and $\mathcal{Z}_x(\delta) = \sum_w \langle \hat{\mathcal{O}}_{\delta,w}(x)\hat{\mathcal{O}}_{\delta,-w}(0)\rangle$. Here $\hat{\mathcal{O}}_{\delta,w}(x) = \hat{O}^+_{\delta,w}(x)\hat{O}^-_{\delta,w}(x)$ are BCCs inserted in the ket and bra at $x$, and the correlator $\langle \dots \rangle_0$ is w.r.t the DBC background with no BCCs inserted. For the two-point functions studied in this work, it is sufficient to average over a *single* pair of BCCs inserted at 0 and $x$ leading to an average over a single parameter $\delta$; additional BCC insertions generically give rise to higher powers and are consequently sub-leading. Additionally, when we write above $\hat{O}_{\delta,w}(x)\hat{A}(x)$, it is to be understood through a point-splitting procedure and a further boundary OPE expansion [11]:

$$\hat{O}_{\delta,w}(x)\hat{A}(x+\epsilon) \underset{\epsilon\to 0}{\sim} \sum_k \epsilon^{\Delta_k - \Delta_{\delta,w} - \Delta_A} C^{\hat{O}_k}_{\hat{O}_{\delta,w}\hat{A}} \hat{O}_k(x), \tag{66}$$

where $\Delta_k, \Delta_{\delta,w}, \Delta_A$ are the boundary scaling dimensions of the operators $\hat{O}_k, \hat{O}_{\delta,w}$, and $\hat{A}$ respectively. Note that the operators $\hat{O}_k$ are in the set of the allowed boundary operators with DBC on the field $\varphi$. We will typically be interested in the operator $\hat{O}_k$ in the above expansion with the smallest scaling dimension since we are interested in the slowest universal decay in (64). Next, since the bra and ket are constrained to have the same measurement outcomes, one must insert identical BCCs in both the bra and the ket, as done in Eq. (65) through the composite operator $\hat{\mathcal{O}}_{\delta,w}$. This is however a specific feature of the TLL measurement problem here. For general quantum critical ground states governed by CFTs, we expect an analog of Eq. (64) to hold but where the boundary renormalizes to an ensemble of genuine CIBCs of the doubled CFT. In that case, the above equations would include genuine BCCs of the doubled theory as well.

With the above prescription in place, let us briefly re-derive the correlators considered in this work, beginning with the (smooth) density correlator $\mathcal{N}_k(x)$ (see Eq. (17)). Since we are now considering a boundary theory with DBC on $\varphi$, non-trivial operators include the dual fields, such as $e^{i\theta}$ and $\partial^n\theta$. Therefore, close to the boundary, we must decompose $\partial_x\varphi$ in terms of these boundary operators. Such a decomposition is precisely given in the bulk-boundary OPE data [12],

$$\partial_x\hat{\varphi} \underset{y\to 0}{\sim} y\partial_x^2\hat{\theta} + \dots, \tag{67}$$

where we consider placing the operator at a small distance $y \sim a$ from the boundary. Here $a$ is the UV cutoff, $\theta$ is the dual boundary field with ellipses representing operators of higher scaling dimensions. Therefore, we must insert $\partial_x^2\theta$ when using Eq. (65) for the smooth density moments. Equivalently, near the boundary, the smooth density is an operator of scaling dimension 2. This already gives us the fixed point contribution to the smooth density correlator [2]

$$\langle \partial_x\varphi(x_1)\partial_x\varphi(0)\rangle_0 \underset{y\to 0}{\sim} y^2\langle \partial_x^2\hat{\theta}(x_1)\partial_x^2\hat{\theta}(0)\rangle_0 \tag{68}$$

$$\sim \frac{y^2}{x^4}. \tag{69}$$

Furthermore, since the one point function $\langle \partial_x^2\hat{\theta}\rangle_0 = 0$, the above result implies that the fixed point contribution to the $k$th moment of the averaged smooth density $\mathcal{N}_k(x)$ is trivially given as $x^{-4k}$.

Next, in order to compute the more interesting contribution to $\mathcal{N}_k(x)$ due to BCCs, we will need the boundary OPE relation for the smooth density with a phase slip of charge $q$:

$$\partial_x^2\hat{\theta}(x)e^{iq\theta(x)} \sim -qe^{iq\theta(x)} + \dots, \tag{70}$$

where we have now also omitted short distance nonuniversal factors and will continue to do so henceforth since we are only interested in the universal long-distance contributions. Crucially, we see that the OPE of the smooth density at the boundary with the BCC $e^{iq\theta}$ returns a factor proportional to the charge $q$ of the BCC. This is a

restatement of a fact that we have already seen in the replica calculation: in the two-slip sector with the solution $\varphi(y) = \pi U_{Q-1}\Delta\boldsymbol{w}[\theta(y-x) - \theta(y)]$ (see Sec. II A), $\partial_y\varphi \propto U_{Q-1}\Delta\boldsymbol{w}[\delta(y-x) - \delta(y)]$. Using Eq. (70), we thus obtain

$$\begin{aligned}\langle\hat{\mathcal{O}}_{\delta,w}(0)\partial_x\hat{\varphi}(0)\partial_x\hat{\varphi}(x)\hat{\mathcal{O}}_{\delta,-w}(x)\rangle_0 &= \langle\hat{O}^+_{\delta,w}(0)\partial_x\hat{\varphi}(0)\hat{O}^+_{\delta,-w}(x)\partial_x\hat{\varphi}(x)\rangle_0\langle\hat{O}^-_{\delta,-w}(x)\hat{O}^-_{\delta,w}(0)\rangle \\ &\sim \left(w+\frac{\delta}{\pi}\right)^2\langle\hat{O}^+_{\delta,-w}(x)\hat{O}^+_{\delta,w}(0)\rangle_0\langle\hat{O}^-_{\delta,-w}(x)\hat{O}^-_{\delta,w}(0)\rangle_0 \\ &\sim \left(w+\frac{\delta}{\pi}\right)^2 x^{-\frac{2}{K}(w+\frac{\delta}{\pi})^2},\end{aligned} \tag{71}$$

and similarly for the correlator(s) $\langle\hat{\mathcal{O}}_{\delta,v}(0)\partial\hat{\varphi}(0)\hat{\mathcal{O}}_{\delta,v}(x)\rangle_0$, where one pulls down only a factor linear in the charge $(w + (\delta/\pi))$. In the first equation, we have factorized the BCCs in the bra and the ket and the density $\partial_x\hat{\varphi}$ acts on the bra. One could have instead acted it on the ket and it would give the same result. In the last line, we used the OPE relation

$$e^{iq\theta(x)}\, e^{iq'\theta(x)} \sim e^{i(q+q')\theta(x)} + \ldots, \tag{72}$$

where we have only kept the leading operator. Finally, using these OPE relations in Eq. (64) with $\hat{A} = -\partial_x\hat{\varphi}$, we have

$$\mathcal{N}_k(x) = \sqrt{\frac{2\log x}{K\pi}}\int_0^\pi \frac{d\delta}{\pi}\,\mathcal{Z}_x(\delta)\left[\frac{\sum_{v\in\mathbb{Z}}\left(v+\frac{\delta}{\pi}\right)^2 x^{\left[-\frac{2}{K}\left(v+\frac{\delta}{\pi}\right)^2\right]}}{\mathcal{Z}_x(\delta)} - \left(\frac{\sum_{v\in\mathbb{Z}}\left(v+\frac{\delta}{\pi}\right)x^{\left[-\frac{2}{K}\left(v+\frac{\delta}{\pi}\right)^2\right]}}{\mathcal{Z}_x(\delta)}\right)^2\right]^k, \tag{73}$$

where we have omitted an overall sign factor, and where $\mathcal{Z}_x(\delta) \equiv C_x(\delta,\mathbb{I}) = \sum_{w\in\mathbb{Z}} x^{-\frac{2}{K}(w+\frac{\delta}{\pi})^2}$ is the same as used in the main text. Up to nonuniversal contributions, the above is the same as Eq. (38) that was obtained via the replica calculation.

Next, for the phase correlators $\Theta_k$ (see Eq. (45)), the bulk operator $e^{i\theta}$ with scaling dimension $1/(4K)$ decomposes into a boundary operator $e^{i\theta}$ with scaling dimension $1/(2K)$ near the boundary [12]. We already used this fact in Sec. III A where we used it to evaluate correlators in the replicated theory and obtain the averaged correlators using the replica limit. Here we will use it directly in the averaging prescription (65) to obtain the leading universal scaling of $\Theta_k(x)$. The fixed point contribution is once again trivially obtained: since the boundary scaling dimension of $e^{i\theta}$ is $1/(2K)$ and the one point function $\langle e^{i\theta(x)}\rangle_0 = 0$, the fixed point contribution to $\Theta_k \sim (x^{-1/K})^{2k}$, as also obtained with the replica calculation (see Eq. (54)). For the contribution to $\Theta_k$ due to leading order BCC insertions, we must adapt Eq. (65) since $\Theta_k$ is not the $k$-th moment of an ordinary two point correlator: it involves $k$ moments of the post-measurement correlator $e^{i[\theta(x)-\theta(0)]}$ and $k$ moments of its complex conjugate. The averaging prescription then takes the form:

$$\Theta_k(x) \sim \frac{\int_0^\pi d\delta\;\mathcal{Z}_x(\delta)\;[\Theta_x(\delta)]^k}{\int_0^\pi d\delta\;\mathcal{Z}_x(\delta)}, \tag{74}$$

where,

$$\Theta_x(\delta) = \frac{1}{\mathcal{Z}_x(\delta)^2}\left(\sum_{w\in\mathbb{Z}}\langle\hat{\mathcal{O}}_{\delta,w}(0)e^{i[\theta(x)-\theta(0)]}\hat{\mathcal{O}}_{\delta,-w}(x)\rangle_0\right)\left(\sum_{v\in\mathbb{Z}}\langle\hat{\mathcal{O}}_{\delta,v}(0)e^{-i[\theta(x)-\theta(0)]}\hat{\mathcal{O}}_{\delta,-v}(x)\rangle_0\right). \tag{75}$$

The above naturally follows from the claim that an average over measurement outcomes is effectively replaced in the IR by an average over BCC insertions appropriately weighted. We can compute one of the factors in the above by using the OPE relation (72) for two phase slip correlators, giving

$$\langle\hat{\mathcal{O}}_{\delta,w}(0)e^{i[\theta(x)-\theta(0)]}\hat{\mathcal{O}}_{\delta,-w}(x)\rangle_0 = \langle\hat{O}^+_{\delta,w}(0)e^{-i\theta(0)}\hat{O}^+_{\delta,-w}(x)e^{i\theta(x)}\rangle_0\langle\hat{O}^-_{\delta,-w}(x)\hat{O}^-_{\delta,w}(0)\rangle_0 \tag{76}$$

$$\sim x^{-\frac{1}{K}\left[\left(\frac{\delta}{\pi}+w-1\right)^2+\left(\frac{\delta}{\pi}+w\right)^2\right]} = x^{-\frac{2}{K}\left[\left(w+\frac{\delta}{\pi}-\frac{1}{2}\right)^2\right]-\frac{1}{2K}} \tag{77}$$

where the first line factorizes the BCCs in the bra and ket and the phase operator are included in the bra, and the second uses the OPE (72) and evaluates the resulting correlators. One can check that the other factor in (75) (with $e^{-i[\theta(x)-\theta(0)]}$) returns $x^{-\frac{2}{K}\left[\left(v+\frac{\delta}{\pi}+\frac{1}{2}\right)^2\right]-\frac{1}{2K}}$. But note that summing over the respective integers $v, w \in \mathbb{Z}$ these two factors are the same:

$$\sum_{w\in\mathbb{Z}} x^{-\frac{2}{K}\left[\left(w+\frac{\delta}{\pi}-\frac{1}{2}\right)^2\right]-\frac{1}{2K}} = x^{-\frac{1}{2K}}\sum_{v\in\mathbb{Z}} x^{-\frac{2}{K}\left[\left(v+\frac{\delta}{\pi}+\frac{1}{2}\right)^2\right]} = x^{-\frac{1}{2K}}\mathcal{Z}_x(\delta+\pi/2), \tag{78}$$

due to the invariance of the sum under $w \to w \pm 1$. This gives

$$\Theta_x(\delta) = \left[ \frac{x^{-\frac{1}{2K}} \mathcal{Z}_x(\delta + (\pi/2))}{\mathcal{Z}_x(\delta)} \right]^2 , \tag{79}$$

and consequently,

$$\Theta_k(x) \sim \sqrt{\frac{2\log x}{K\pi}} \int_0^\pi \frac{d\delta}{\pi} \mathcal{Z}_x(\delta) \left[ \frac{x^{-\frac{1}{2K}} \mathcal{Z}_x(\delta + (\pi/2))}{\mathcal{Z}_x(\delta)} \right]^{2k} , \tag{80}$$

which is the same as Eq. (60) (up to nonuniversal additive and multiplicative terms) that we obtained using the replica calculation.

Although we have explicitly verified this general picture only for the two correlators considered in this work, the same prescription can, in principle, be used to compute arbitrary two-point correlation functions in the TLL. More broadly, as discussed above and in the main text, we expect a closely related physical picture to apply to other quantum-critical ground states governed by CFTs, although with important differences from the TLL case. When measurements constitute a relevant boundary perturbation, and provided that the RG flow does not terminate at an intermediate finite-measurement fixed point such as those identified in Ref. [13], they should drive the doubled CFT toward an ensemble of conformally invariant boundary conditions. The resulting long-distance physics is then determined by the boundary operator content and spectrum associated with this ensemble.

For the TLL, the structure of these infrared boundary conditions is particularly tractable. The relevant ensemble consists entirely of DBCs of a ($c = 1$) compact free-boson CFT, whose boundary states, boundary operators, and BCC spectra are well understood. In a generic CFT, however, identifying the conformally invariant boundary conditions selected by the measurement-induced RG flow may itself be a difficult problem. Moreover, the relevant boundary conditions and BCC operators belong to the *doubled* theory, and their structure need not factorize between the bra and the ket. Determining this boundary data, together with the appropriate average over the resulting ensemble, may therefore be substantially more challenging beyond the free-boson setting.

## V. NUMERICS

We test our predictions through tensor-network calculations using iTensor [14], for a spin-1/2 lattice model governed by the Hamiltonian,

$$H = \sum_i \left[ -t(S_i^+ S_{i+1}^- + S_i^- S_{i+1}^+) + U_1 S_i^z S_{i+1}^z + U_2 S_i^z S_{i+2}^z \right] , \tag{81}$$

where $S_i^\pm = S_i^x \pm i S_i^y$ and we consider periodic boundary conditions system sizes $L = 84, 120$. The model reduces to free fermions when $U_1 = U_2 = 0$. Away from this limit however, the above model is generally nonintegrable and its ground state can realize a variety of phases, including the standard Luttinger liquid and, notably, a pair-cluster Luttinger liquid [15, 16]. Importantly, the next-nearest-neighbor interaction allows us to access a standard Luttinger-liquid regime with $K < 1/2$, for which measurements are relevant. We benchmark our predictions using two ground states with distinct values of $K < 1/2$. Approximate ground states are obtained using DMRG [17, 18] at fixed filling fraction $n = 1/3$, with $t = 1$ and interaction parameters $(U_1, U_2) = (10, 1.0)$ and $(10, 2.5)$. The bond dimensions for $L = 84$ are $(376, 404)$ and for $L = 120$ are $(507, 544)$ for the parameters $(U_1, U_2) = (10, 1.0)$ and $(U_1, U_2) = (10, 2.5)$ respectively.

### A. Extracting the Luttinger Parameter

Since there is no analytic relation between $(U_1, U_2)$ and $K$, we extract the Luttinger parameter from the algebraically decaying $K$-dependent correlation functions in the ground state. In particular, we use [1]

$$\langle \hat{X}_j \hat{X}_{j+r} \rangle_{0,c} \sim A_0 d_c(r)^{-1/(2K)} + A_1 \cos\left(\frac{2\pi r}{3}\right) d_c(r)^{-\frac{1}{2K}+2K}, \tag{82}$$

$$\langle \hat{Z}_j \hat{Z}_{j+r} \rangle_{0,c} \sim -\frac{K}{2\pi^2} \frac{1}{d_c(r)^2} + A_2 \cos\left(\frac{2\pi r}{3}\right) d_c(r)^{-2K}, \tag{83}$$

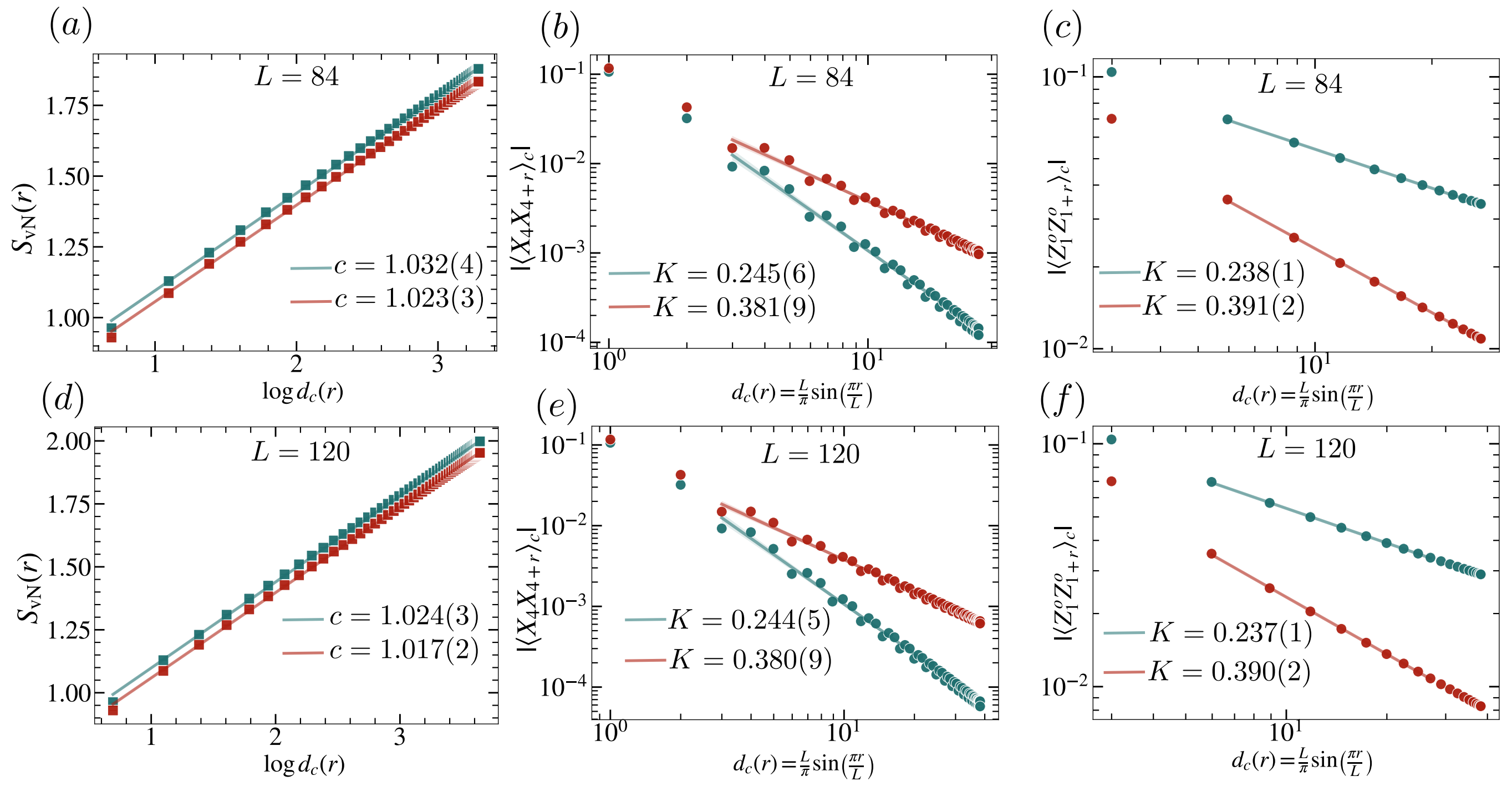


FIG. 1: **Extracting the central charge $c$ and Luttinger parameter $K$.** Green markers show data for the ground state at $(U_1, U_2) = (10, 2.5)$, while red markers correspond to $(U_1, U_2) = (10, 1.0)$. Fits are shown as solid lines of the corresponding color with reduced opacity. (a,d) Von Neumann entanglement entropy $S_{\mathrm{vN}}(r)$ plotted against $\log d_c(r)$ for $L = 84$ and $L = 120$, respectively. The central charge is inferred from the slope $c/3$ from Eq. (84). (b,e) Connected correlator $|\langle X_4 X_{4+r}\rangle_c|$ plotted against $d_c(r)$ and fitted to $d_c(r)^{-1/(2K)}$ for $L = 84$ and $L = 120$, respectively. (c,f) Correlator $\langle Z_1^o Z_{1+r}^o\rangle$ plotted against $d_c(r)$, and fitted to $d_c(r)^{-2K}$ for $L = 84$ and $L = 120$, respectively.

where $\hat{X}_j = 2S_j^x$, $\hat{Z}_j = 2S_j^z$, $d_c(r) = (L/\pi)\sin(\pi r/L)$ is the appropriate chord length for a finite ring of length $L$, and $\langle \ldots \rangle_{0,c}$ denotes the connected correlator in the ground state. For the two-point $\hat{X}$ correlator, the asymptotic decay for $K < 1/2$ is $\langle \hat{X}_j \hat{X}_{j+r}\rangle_0 \sim d_c(r)^{-1/(2K)}$, which allows us to extract $K$, as shown in Fig. 1. We use the two-point $\hat{Z}$ correlator as an additional cross-check of the extracted value. Since the smooth contribution in Eq. (83) decays as $d_c(r)^{-2}$ and does not provide a direct estimate of $K$ through its exponent, we instead consider the operator $\hat{Z}_j^o = (2\hat{Z}_{j-2} - \hat{Z}_{j-1} - \hat{Z}_j)/3$. This combination filters out the smooth contribution while retaining the oscillating term with wave vector $2\pi/3$, which decays as $d_c(r)^{-2K}$; see Fig. 1. For each parameter set, the values extracted from the two correlators agree closely, and we use their average. This gives $K \approx 0.385$ for $(U_1, U_2) = (10, 1.0)$ and $K \approx 0.241$ for $(U_1, U_2) = (10, 2.5)$. Finally, as an additional consistency check, we verify the ground-state entanglement scaling using the Calabrese-Cardy form [19, 20],

$$S_{\mathrm{vN}}(r) \sim \frac{c}{3}\log d_c(r), \tag{84}$$

where $S_{\mathrm{vN}}(r)$ is the von Neumann entanglement entropy of a subsystem of size $r$. By computing $S_{\mathrm{vN}}(r)$ for several subsystem sizes and fitting to the above form, we extract $c \approx 1$, as expected for a Luttinger liquid (see Fig. 1).

### B. Measurement-averaged correlators

After obtaining the approximate ground states, we perform local density measurements with outcomes sampled from the Born distribution. This is done as follows: at each site $i$, we apply the on-site measurement operator $\hat{M}_{m_i} = [2\cosh(2\gamma)]^{-1/2} e^{\gamma m_i \hat{Z}_i}$, where $\hat{Z}_i = \mathrm{diag}(1, -1)$ and $m_i = \pm 1$ for all $i \in 1, \ldots, L$. Denoting the normalized

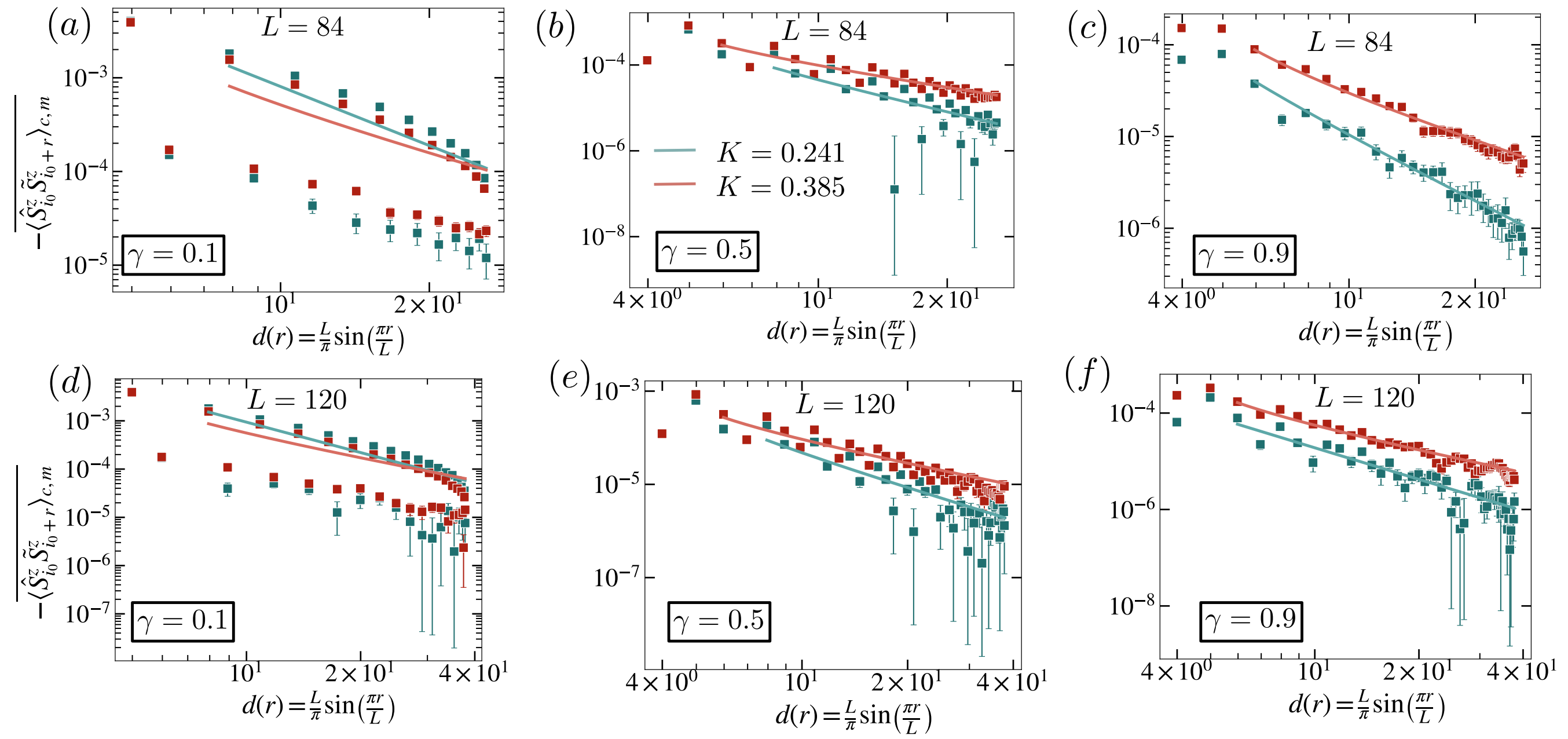


FIG. 2: **Scaling of measurement-averaged correlator** $-\sum_m p_m \langle \hat{S}^z_{i_0} \tilde{S}^z_{i_0+r} \rangle_{m,c}$ **vs chord length** $d(r)$ **for** $L = 84, 120$ **and measurement strengths** $\gamma = 0.1, 0.5, 0.9$**.** Green and red markers correspond to ground states with Luttinger parameters $K = 0.241$ and $K = 0.385$, respectively. Solid lines of the corresponding colors show fits to the theoretical predictions of the connected correlator of smooth density, $\mathcal{N}_1(r)$, made in Eq. (44).

state before measuring site $i$ by $|\psi_{i-1}\rangle$, we evaluate the conditional Born probabilities $p_{m_i} = \langle \psi_{i-1}|\hat{M}^2_{m_i}|\psi_{i-1}\rangle$, which satisfy $p_{+1} + p_{-1} = 1$. We then draw a random number $r$ from the uniform distribution $\mathcal{U}[0,1]$ and choose $m_i = +1$ if $r < p_{+1}$ and $m_i = -1$ otherwise. Following this outcome, the state is updated according to $|\psi_i\rangle = \hat{M}_{m_i}|\psi_{i-1}\rangle/\sqrt{p_{m_i}}$. This procedure is repeated sequentially for all sites, starting from the normalized ground state $|\psi_0\rangle$, with $\langle\psi_0|\psi_0\rangle = 1$. For each sampled record $m$, we evaluate the connected two-point correlators with one end fixed at a reference site $i_0$, namely $C^m_z(r) = \langle \hat{S}^z_{i_0} \hat{S}^z_{i_0+r}\rangle_{m,c}$ and $C^m_x(r) = \langle \hat{S}^x_{i_0} \hat{S}^x_{i_0+r}\rangle_{m,c}$, where $\langle \cdots \rangle_{m,c}$ denotes the connected expectation value in $|\psi_m\rangle$, e.g. $\langle \hat{S}^z_{i_0} \hat{S}^z_{i_0+r}\rangle_{m,c} = \langle \hat{S}^z_{i_0} \hat{S}^z_{i_0+r}\rangle_m - \langle \hat{S}^z_{i_0}\rangle_m \langle \hat{S}^z_{i_0+r}\rangle_m$. Born-averaged moments are then estimated through a mean over $N_{\text{samples}}$ independently drawn records $m^{(1)}, \ldots, m^{(N_{\text{samples}})}$, that is, $\overline{[C_x(r)]^k} \equiv \sum_m p_m\, [C^m_x(r)]^k = \lim_{N_{\text{samples}}\to\infty} N^{-1}_{\text{sampls}} \sum_{n=1}^{N_{\text{samples}}} [C^{m^{(n)}}_x(r)]^k$, and analogously for $C_z$; since the records are sampled from $p_m$, no reweighting is required. Finally, the samples $C^m_{z,x}(r)$ must be post-processed to obtain the correlators $\mathcal{N}_k(r)$ and $\Theta_k(r)$, as we now describe.

Using Bosonization [1], we have

$$S^z_j \simeq -\frac{1}{6} - \pi^{-1}\partial_x\varphi + A_1 \cos(2k_F x_j - 2\varphi(x_j)), \tag{85}$$

$$S^x_j \simeq \cos\theta(x_j)[B_0 + B_1 \cos(2k_F x_j - 2\varphi(x_j))], \tag{86}$$

where $x_j = d_c(ja)$, where $a$ is the UV cutoff and $d_c(ja)$ is the chord length. First, since $\mathcal{N}_k(r)$ is the $k$th moment of the connected two point correlator of the *smooth* part of density, we must remove the leading oscillatory component in the correlator samples $C^m_z(r)$ that stems from the $2k_F$ component of density (see Eq. (2)). We do so by introducing the smoothening operator $\tilde{S}^z_j = (\hat{S}^z_{j-1} + \hat{S}^z_j + \hat{S}^z_{j+1})/3$ and evaluate the correlator

$$\sum_m p_m \langle \hat{S}^z_{i_0} \tilde{S}^z_{i_0+r}\rangle^k_{m,c} = \overline{[C_{\tilde{z}}(r)]^k} \sim \sum_m p_m \langle \partial\varphi(x_{i_0})\partial\varphi(x_{i_0+r})\rangle^k_{m,c}, \tag{87}$$

where $\overline{[C_{\tilde{z}}(r)]} = \overline{[C_z(r-1)] + [C_z(r)] + [C_z(r+1)]}/3$, and the above follows from Eq. (85) (up to factors). The plots for $k = 1, 2$ are shown in Figs. 2 and 3 for system sizes $L = 84, 120$ and $\gamma = 0.1, 0.5, 0.9$, where we fit the asymptotic decay $e^{-\chi/2}/\sqrt{\chi}$ with $\chi = K^{-1}\log d_c(r) + \kappa(\gamma)$, and where $\kappa(\gamma)$ and a non-universal pre-factor are the fitting parameters.

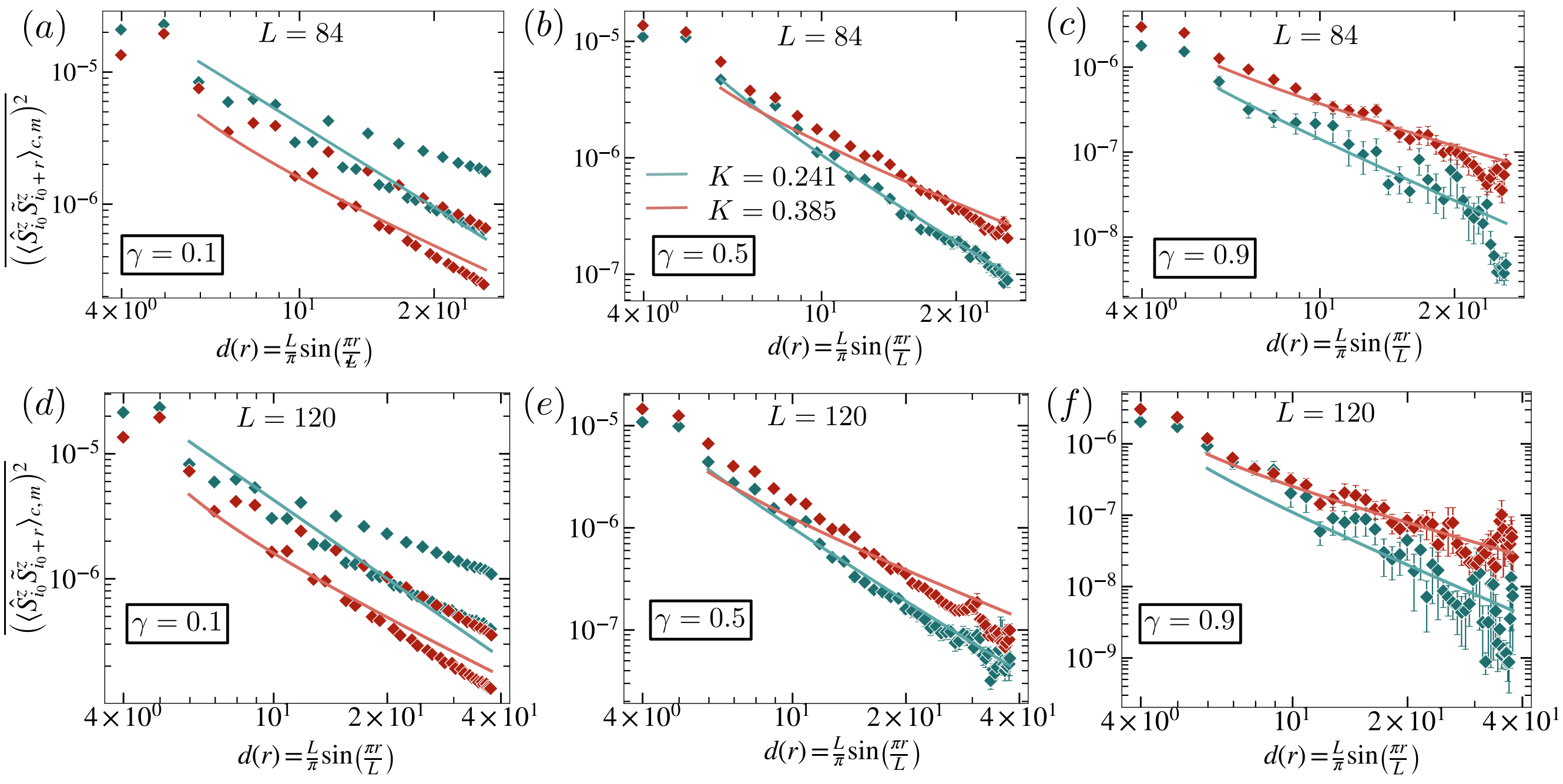


FIG. 3: **Scaling of measurement-averaged correlator $\sum_m p_m \langle \hat{S}^z_{i_0} \tilde{S}^z_{i_0+r}\rangle^2_{m,c}$ vs chord length $d(r)$ for $L = 84, 120$ and measurement strengths $\gamma = 0.1, 0.5, 0.9$.** Green and red markers correspond to ground states with Luttinger parameters $K = 0.241$ and $K = 0.385$, respectively. Solid lines of the corresponding colors show fits to the theoretical predictions of the connected correlator $\mathcal{N}_2(r)$ made in Eq. (44).

For $\Theta_k(r)$, we consider $k = 1$ and similarly plot

$$\sum_m p_m \langle \hat{S}^x_{i_0} \tilde{S}^x_{i_0+r}\rangle^{2k}_{m,c} = \overline{[C_{\tilde{x}}(r)]^{2k}} \simeq \sum_m p_m |\langle e^{i[\theta(x_{i_0+r})-\theta(x_{i_0})]}\rangle|^{2k}_{m,c}, \tag{88}$$

where $\tilde{S}^x_j = (\hat{S}^x_{j-1} + \hat{S}^x_j + \hat{S}^x_{j+1})/3$ and $\overline{[C_{\tilde{x}}(r)]} = \overline{[C_x(r-1)] + [C_x(r)] + [C_x(r+1)]}/3$ removes the sub-leading oscillatory component as seen from Eq. (86). The above equation also uses the fact that density measurement preserve the $U(1)$ symmetry and so charged two point correlators of the form $\langle S^{\pm}_j S^{\pm}_{j+r}\rangle_m$ and charges one point functions $\langle S^{\pm}_j\rangle_m$ vanish identically, and that the correlator $\langle e^{i[\theta(x_{i_0+r})-\theta(x_{i_0})]}\rangle_m$ is real for all measurement outcomes. The plots for $k = 1$ are shown in Fig. 4 for system sizes $L = 84, 120$ and $\gamma = 0.1, 0.5, 0.9$, where we fit the asymptotic decay $e^{-\chi/2}/\sqrt{\chi}$ with $\chi = K^{-1} \log d_c(r) + \kappa(\gamma)$, and where $\kappa(\gamma)$ and a non-universal pre-factor are the fitting parameters.

As expected, the quality of the fits improves systematically with increasing measurement strength: for the weak measurement strength $\gamma = 0.1$, the fits are poor, since at this coupling the system has not yet crossed over to the strong-coupling fixed point about which our analytical results are derived, while the agreement improves at $\gamma = 0.5$ and improves further at $\gamma = 0.9$. Working at even larger $\gamma$ does not improve matters further in practice: increasing the measurement strength suppresses the connected correlations by several orders of magnitude, so the corresponding signal is increasingly noisy at accessible sample sizes. Intermediate-to-strong measurement strengths of $\gamma \sim 0.7$–$0.9$ therefore offer the best window, and there we find good agreement between the numerical data and the predicted asymptotic decay.

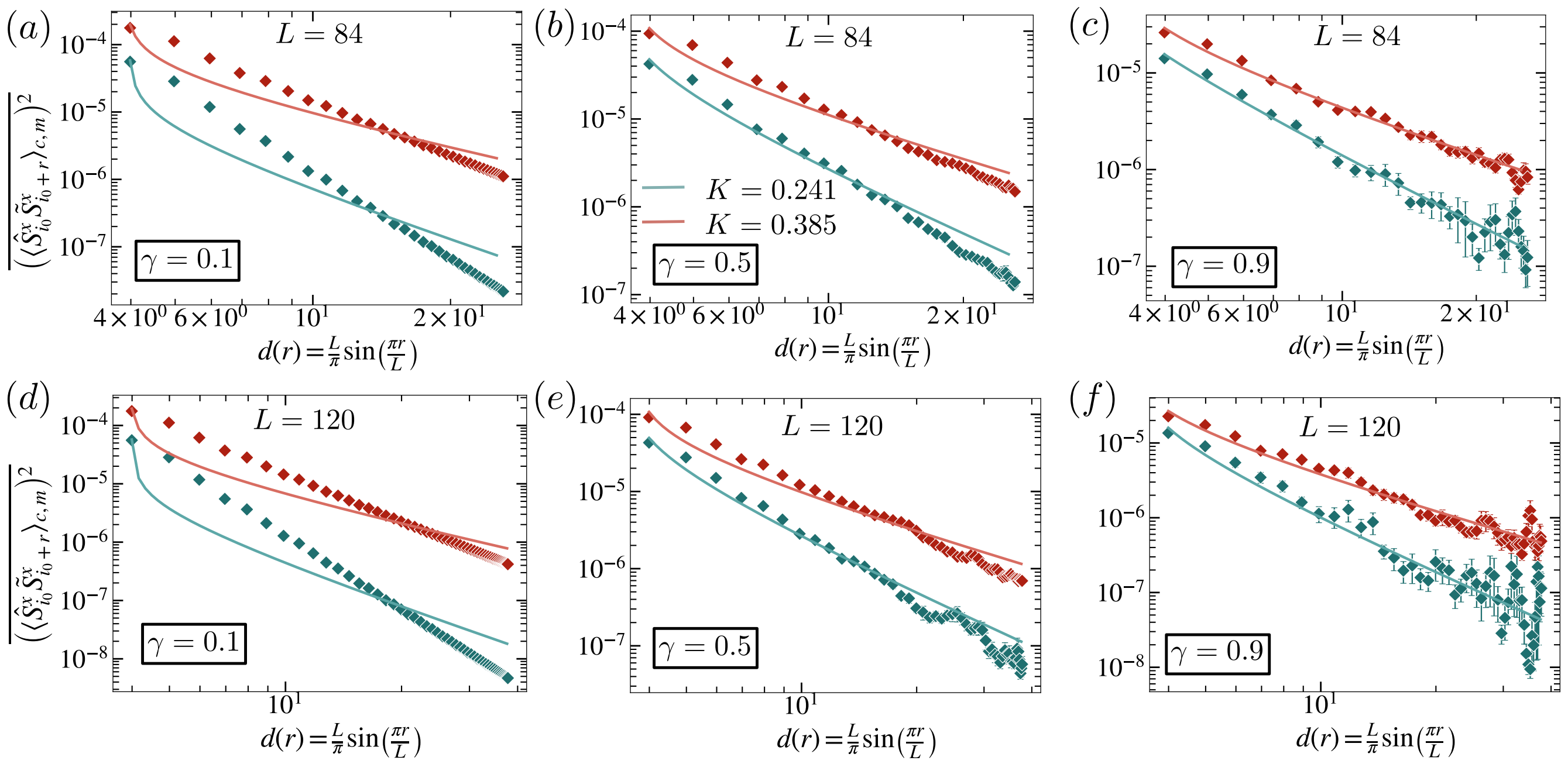


FIG. 4: **Scaling of measurement-averaged correlator $\sum_m p_m \langle \hat{S}^x_{i_0} \tilde{S}^x_{i_0+r}\rangle^2_{m,c}$ vs chord length $d(r)$ for $L = 84, 120$ and measurement strengths $\gamma = 0.1, 0.5, 0.9$.** Green and red markers correspond to ground states with Luttinger parameters $K = 0.241$ and $K = 0.385$, respectively. Solid lines of the corresponding colors show fits to the theoretical predictions of the connected correlator $\Theta_1(r)$, made in Eq. (61).


\end{document}